\documentclass[12pt,preprint]{aastex}

\slugcomment{\bf Accepted August 27, 2003 to AJ}

\shorttitle{A Spectroscopic Technique for Measuring Stellar Properties of Pre--Main Sequence Stars}
\shortauthors{Doppmann \& Jaffe}

\begin{document}

\title{A Spectroscopic Technique for Measuring Stellar Properties of Pre--Main Sequence Stars}

\author{G. W. Doppmann\altaffilmark{1,2,3} and 
D. T. Jaffe\altaffilmark{2,3}}

\email{gdoppmann@mail.arc.nasa.gov}
\email{dtj@astro.as.utexas.edu}

\altaffiltext{1}{NASA Ames Research Center, MS 245-6, Moffett Field, CA 94035-1000}
\altaffiltext{2}{Department of Astronomy, 1 University Station C1400, Austin, TX 78712-1083}
\altaffiltext{3}{Visiting Astronomer, Kitt Peak National Observatory, National Optical Astronomy Observatory, which is operated by the Association of Universities for Research in Astronomy, Inc. (AURA) under cooperative agreement with the National Science Foundation.}

\begin{abstract}

We describe a technique for deriving effective temperatures, surface
gravities, rotation velocities, and radial velocities from high
resolution near--IR spectra.  The technique matches the observed
near--IR spectra to spectra synthesized from model atmospheres. Our
analysis is geared toward characterizing heavily reddened pre--main
sequence stars but the technique also has potential applications in
characterizing main sequence and post--main sequence stars when these
lie behind thick clouds of interstellar dust.  For the pre--main
sequence stars, we use the same matching process to measure the amount
of excess near--IR emission (which may arise in the protostellar
disks) in addition to the other stellar parameters.  The information
derived from high resolution spectra comes from line shapes and the
relative line strengths of closely spaced lines.  The values for the
stellar parameters we derive are therefore independent of those
derived from low resolution spectroscopy and photometry.  The new
method offers the promise of improved accuracy in placing young
stellar objects on evolutionary model tracks.  Tests with an
artificial noisy spectrum with typical stellar parameters, and
signal--to--noise of 50 indicates a 1$\sigma$ error of 100 K in
T$_{\rm eff}$, 2 km s$^{-1}$ in $v\sin i$, and 0.13 in continuum
veiling for an input veiling of 1.  If the line flux ratio between the
sum of the Na, Sc, and Si lines at 2.2 $\mu$m and the (2--0)~$^{12}$CO
bandhead at 2.3 $\mu$m is known to an accuracy of 10\%, the errors in
our best fit value for $\log g$ will be $\Delta \log g$~=~0.1--0.2.
We discuss the possible systematic effects on our determination of the
stellar parameters and evaluate the accuracy of the results derivable
from high resolution spectra. In the context of this evaluation, we
explore quantitatively the degeneracy between temperature and gravity
that has bedeviled efforts to type young stellar objects using low
resolution spectra. The analysis of high resolution near--IR spectra
of MK standards shows that the technique gives very accurate values
for the effective temperature.  The biggest uncertainty in comparing
our results with optical spectral typing of MK standards is in the
spectral type to effective temperature conversion for the standards
themselves.  Even including this uncertainty, the 1$\sigma$ difference
between the optical and IR temperatures for 3000--5800~K dwarfs is
only 140~K.  In a companion paper \citep*{doppmann2002b}, we present
an analysis of heavily extincted young stellar objects in the
$\rho$~Ophiuchi molecular cloud.

\end{abstract}


\section{Introduction}

Our understanding of main sequence and post main sequence stellar
evolution is a triumph of collaboration between theorists building
physical models and observers collecting precise observations.  Over
many decades, the two groups have compared their results on the common
ground of the Hertzsprung--Russell (H--R) diagram.  For young stellar
objects (YSOs) however, the link between theory and observation is
much more tenuous.

There is a well--developed empirical scheme that classifies YSOs by
their broad--band spectral energy distributions \citep{lada1987}.
Class I sources have rising near-- to mid--IR spectra.  Class II
sources have broad, largely flat infrared spectral energy
distributions, and Class III sources have spectral energy
distributions of hot reddened blackbodies.  Since the emission that
distinguishes Class I and II objects from Class III sources does not
arise from the stellar photosphere, models of pre--main sequence (PMS)
evolution are not strongly constrained by matching theoretical and
observed spectral energy distributions. Comparison of model tracks
with observational H--R diagrams has been possible for visible T~Tauri
(Class II) stars (TTSs) \citep[e.g.][]{huang1961,stahler1988,
hartmann1991, kenyon1995, white1999, webb1999, simon2000,
white2001}. Obscuration by dust makes it difficult to study Class I
objects and to observe TTSs in the visible if these objects are in
clusters or associations within molecular clouds.  The resulting
difficulties in comparing tracks with observations leave our
understanding of the physics of YSOs on a much less solid footing than
our understanding of more evolved objects.
 
In studying YSOs, our main goal is to understand the history of these
objects from their formation to their arrival on the main sequence.
We also wish to use the ensemble of such objects to characterize
forming clusters and associations; their initial mass functions and
star formation histories.  To meet these goals, we need to determine
parameters of the stars and relate them to other properties of
protostellar disks and of the surrounding cloud.  The most important
stellar parameters to obtain are the effective temperature and the
luminosity or surface gravity since this pair permits us to place the
YSOs into a theoretical H--R diagram for comparison with theoretical
PMS evolution models.  (Note that luminosity and $\log g$ are not
fully interchangeable since the relationship between these two
quantities depends on the not yet well established
mass--luminosity--radius relationship for YSOs.)  Other useful
parameters include the amount of reddening or extinction, the amount
of non--photospheric emission as a function of wavelength, and the
stellar rotation rate.

Observers have used many techniques to investigate the properties of
stars in very young embedded clusters.  In well--studied regions like
the $\rho$~Ophiuchi cloud core, previous investigators have used
photometric surveys in the near--IR \citep{wilking1983, greene1992,
barsony1997}, in the mid--IR \citep{bontemps2001}, and
low--to--moderate resolution ($R\equiv\lambda/\Delta\lambda
=$~500--2000) near--IR spectroscopy \citep{casali1992,greene1995,
greene1996, kenyon1998, luhman1999}, to estimate temperatures,
luminosities, and the amount of excess (non--photospheric emission) in
the infrared for the sources.  In embedded clusters, however, even
near--IR photometry and low resolution spectroscopy suffer under
disadvantages not inflicted upon these techniques when applied to main
sequence stars or to less heavily extincted young stars. Problems
include extremely high extinction \citep[e.g. the central part of the
$\rho$~Ophiuchi molecular cloud where $A_{\rm v}=40\pm10.9$
magnitudes,][]{luhman1999}, and excess emission in the near-- and
mid--IR \citep[from warm dust in the circumstellar
disks,][]{GWAYL1994,strom1995}.

Observers recognized long ago that spectral classification in the
near--IR was a potentially valuable tool for deriving the properties
of obscured stars in the Galactic Plane and young stars obscured by
their natal clouds \citep{merrill1979}.  Young stars are brighter in
the infrared both because their photospheres tend to be cool and
because it is easier for the stellar infrared emission to penetrate
through the dust within the star forming cloud or along the
line--of--sight.  Also, in TTSs, the ratio of photospheric flux to the
hot continuum that produces the excess emission frequently seen in the
UV \citep[presumably from accretion shocks,][]{gullbring2000} is
higher in the infrared than at shorter wavelengths, permitting better
detections of photospheric lines.

The near--IR spectra of late--type stars contain useful information
about the luminosities of the targets.  The strength of the CO
overtone bands at 2.3 $\mu$m was recognized early--on as a useful
indicator of luminosity, albeit with additional sensitivity to
temperature \citep{baldwin1973, kleinmann1986, lancon1992}.
\citet{ramirez1997} find that an index formed from the equivalent
widths of the strong near--IR lines of neutral calcium and sodium and
the (2--0)~$^{12}$CO bandhead is a luminosity indicator, independent
of temperature, for giants in the range K0 to M6.

Accurate estimates of spectral type are also possible from near--IR
spectra.  \citet{kleinmann1986} calculated equivalent widths of key
features from their K--band spectra of MK standards and derived the
dependence of these equivalent widths on spectral and luminosity
class. \citet{meyer1998} derived line ratio relations from H--band
spectra of MK standards and found that the relations agree with
optical spectral types to within $\pm 2$ subclasses.  They argue that
using line ratios, rather than equivalent widths, makes the T$_{\rm
eff}$ determination less sensitive to the presence of continuum
excesses in PMS objects.  While most of the efforts to derive spectral
types from the IR spectra have focused on empirical equivalent width
or line ratio to T$_{\rm eff}$ relations, there have also been a
number of efforts to match low resolution H and K--band spectra to
synthetic spectra \citep{kirkpatrick1993, ali1995, leggett1996}.
\citet{ali1995} find that they can match the temperatures of dwarfs
with an error of $\pm$350~K using this technique.

All of the studies we have described so far were carried out with
resolving powers below 3000, where the unsaturated photospheric
features are unresolved.  At these resolving powers, not only can the
depths of the lines get quite small, especially in the presence of
excess continuum emission but there can also be problems due to line
blending.  For example, many of these studies use the equivalent width
of the Na features at 2.2~$\mu$m in the determination of spectral type
and luminosity class.  At $R=1,000$, the typical resolving power for
the studies, however, the Na lines are blended with weaker but
significant lines of Sc and Si that have very different dependences on
T$_{\rm eff}$ and $\log g$.  Existing spectra of the Na interval at a
greater resolving power have shown that the relative strengths of the
Sc, Si, and Na lines plus the shape of the Na features could
potentially be sensitive indicators of effective temperature for
dwarfs \citep{wallace1996, greene1997}.

\begin{figure}
  \epsscale{0.8} \plotone{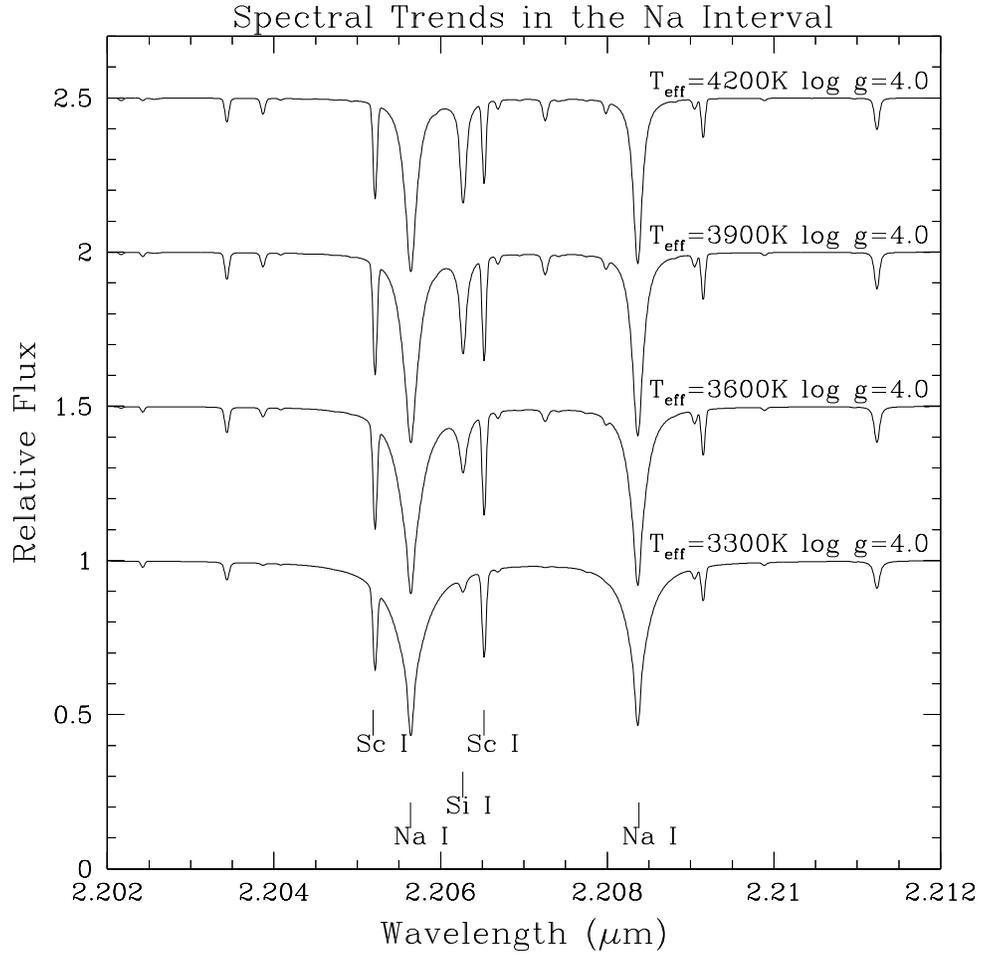}
\caption[K band absorption line behavior with changes in the T$_{\rm eff}$ 
of the synthesis models] 
{\label{fig-na.behavior} A grid of spectral syntheses based on NextGen
\citep{hauschildt1999} atmosphere models.  These spectra illustrate
the change in Na line shape and depth and the variations in Si/Sc for
temperatures from 3300~K to 4200~K for $\log g$ fixed at 4.0.  Spectra
have been smoothed to a R = 50,000.}
\end{figure}

We present here a technique for deriving some of the properties of PMS
stars from high resolution near--IR spectra. High resolution spectra
are particularly useful because the closely spaced lines (like the Na,
Sc, and Si features at 2.206 $\mu$m, see Figures
\ref{fig-lambda.intervals} \& \ref{fig-na.behavior}) in late--type
stars are observable separately and because it is possible to use the
spectra to glean information from the shapes of the absorption lines
\citep{johnskrull2001} The stellar parameters we derive include the
effective temperature (T$_{\rm eff}$), the surface gravity ($\log g$),
the rotation rate ($v\sin i$), and the amount of stellar/circumstellar
emission in excess of the photospheric emission (r$_{\lambda}$). The
method is largely independent of photometric data (except as used to
cross--calibrate high resolution spectral segments at different
wavelengths and to estimate the small amount of reddening difference
between 2.2 $\mu$m and 2.3$\mu$m, see Appendix A) and offers the
promise of improved accuracy, especially for derivations of the
properties of heavily obscured stars.

In this paper, we present a quantitative derivation of the physical
properties of PMS stars from high resolution spectra taken in the
K--band (2.0--2.4~$\mu$m).  Stars in the mass range from 0.1 to
0.9~M$_{\odot}$ will lie in the spectral type range from M6 to K0 from
the time they become visible in the near--IR until they reach the main
sequence \citep{dantona1997,baraffe1998,palla1999,siess2000}.  For
such late--type stars and even for somewhat earlier types (later than
G5), the near--IR spectra are rich in lines of neutral metals and
hardy molecules.  It is therefore possible to use high resolution
spectra from a limited spectral range to provide us with many of the
important physical parameters of young or obscured stars.  Our
analysis involves a comparison of synthetic spectra to the observed
high resolution data.  The observations and data reduction are
described in $\S$~2.  In $\S$~3, we detail the basic spectral analysis
technique with particular emphasis on the features of our method
necessary to deal with the peculiarities of PMS objects.  In $\S$~4,
we analyze the internal errors and investigate inherent systematic
uncertainties.  We compare results for standard stars from optical
spectroscopy to the results we obtain through the analysis of high
resolution near--IR spectra, in $\S$~5, and discuss the ways in which
our derived properties supplement or improve upon properties measured
with other techniques.  In a companion paper \citep*[][hereafter
DJW03] {doppmann2002b}, we present and analyze near--IR spectra of a
sample of Class II YSOs in the core of the $\rho$~Ophiuchi molecular
cloud.

\section{Observations and Data Reduction}\label{sec-observations}
 
In order to test our technique for derivation of stellar parameters
from high resolution spectra in the 2.2 $\mu$m atmospheric window, we
have assembled a sample of spectra of MK standards. Table 1 lists the
stars in this sample, the instruments used to obtain the data, and
various stellar properties obtained from the literature.
 
We observed part of the sample using the PHOENIX spectrograph
\citep{hinkle1998}, on the Kitt Peak 4 meter in May 2000.  The
observed spectral interval covered 95~${\rm \AA}$ centered at 2.2070
$\mu$m.  The resolving power for these observations was
R~$\equiv$~$\lambda$/$\Delta \lambda \cong$ 50,000. Individual pixels
covered a range in wavelength of $\lambda$/$\Delta \lambda$ = 240,000.
 
A partially overlapping set of MK standards was observed using the
NIRSPEC instrument \citep{mclean1998}, on the Keck Telescope in May
2000.  These spectra were kindly provided to us by Tom Greene.  For
these observations, NIRSPEC had a slit--limited resolving power of
17,500 and covered a total of 230 nm over 6 non--contiguous orders in
the 2 $\mu$m atmospheric window.  Observations with both instruments
were made by nodding the telescope, placing the target star
alternately at two different positions along the slit.

IRAF was used to reduce the spectra from both spectrometers in roughly
same way \citep[for details about the NIRSPEC data reduction
see][]{greene2000}.  With the PHOENIX data, we differenced the source
frames taken at alternate slit positions and divided by flat fields
using an internal continuum lamp that uniformly illuminated the entire
slit.  At locations of bad pixels, we substituted values interpolated
from neighboring positions.  We then optimally extracted the spectra
using IRAF's {\tt apall} package.  We used telluric absorption lines
of H$_2$O and CH$_4$ to wavelength calibrate the PHOENIX data. The
best fit to the low order wavelength solution was accurate to 0.1
pixels along the detector array.  We removed the telluric lines from
the spectra by dividing the data by spectra of early--type stars taken
at the same airmass (typically within 5$^{\rm o}$ of the target).  We
produced the final spectra by taking a signal--weighted average of the
calibrated spectra from the two beam positions.  To prepare the final
spectra for comparison with models, we set a continuum level by eye
using regions in the observed spectra where synthesis models indicate
that strong lines and extended line wings are not present.

\section{Method}\label{sec-method}

Ideally, we would like to be able to derive stellar parameters across
the whole range of masses and ages present in clusters or associations
of newly forming stars.  The actual range of surface gravities and
effective temperatures for which we can derive parameters from high
resolution spectroscopic observations using the technique we outline
here is constrained in several ways: we are limited to objects that
permit observable amounts of near--IR radiation from the stellar
photosphere to escape through the surrounding disk and envelope.  In
general, photometric studies of YSOs imply that the lowest surface
gravities for which objects become visible in the near--IR are $\log
g\approx 3.0$ \citep{comeron1993, siess2000}.  Our technique requires
that the stellar photospheres have a sufficient number of reasonably
strong lines and that these lines be sensitive to variations in
temperature and surface gravity.  Further, it requires that the
stellar models and available line lists be adequate to permit us to
make accurate high resolution spectral syntheses for comparison with
the observed spectra.  With the stellar atmospheres and synthesis
program we are using, our analysis technique works well within the
range $3000\leq T_{\rm eff} \leq 5800~{\rm K}$, and $3.5 \leq \log g
\leq 5.5$.  Since H--R diagram evolutionary tracks for low mass PMS
objects are largely vertical, the temperature constraint implies a
range of masses for which we can position objects in the H--R diagram
of roughly 0.1 to 1.6~$M_{\odot}$.

\subsection{Technique Overview}\label{sec-technique}

The basis for our technique is a grid of synthetic high resolution
spectra in the K window (2.0--2.4~$\mu$m).  The grid spans the
relevant ranges of the important stellar parameters for YSOs:
effective temperature, surface gravity, veiling, and $v\sin i$
rotation.  The best model fit is chosen by an RMS minimization of the
residuals across the photospheric absorption lines in our spectral
window. This minimization also includes a fit for the stellar radial
velocity.

The 2.0--2.4~$\mu$m atmospheric window contains features that are
sensitive to both the temperature and pressure in the stellar
photosphere.  It is also the longest wavelength band where the
photospheric emission from the youngest stars is comparable in flux to
the thermal emission from dust in the circumstellar disk.  At this
wavelength, the sensitivity of ground--based spectrometers with large
resolving powers is not yet compromised by thermal emission from the
telescope and sky.  The 2 $\mu$m band is also not far from the maximum
in the photospheric emission from late--type stars.

For the purposes of our spectral matching program, high spectral
resolution means sufficient resolution to permit us to resolve most
stellar lines (R~$\geq$~20,000). No existing high resolution
spectrometer covers the entire 2.0--2.4~$\mu$m atmospheric window with
a single exposure.  The instrument with the most coverage
\citep[NIRSPEC,][]{mclean1998} gives cross--dispersed spectra covering
about 1/3 of the window in disconnected segments.  Other existing
instruments cover only individual 50--200 Angstrom bands
\citep[PHOENIX,][]{hinkle1998}, \citep[CSHELL,][]{greene1993},
\citep[CGS4,][]{mountain1990,wright1993}. In order to be generally
applicable without enormous expenditure of telescope time, our
technique should therefore use only a limited part of the spectrum
available within the atmospheric window.

We have used spectral synthesis models to explore the K window to find
the spectral intervals that have strong lines that vary significantly
with variations in the stellar parameters.  Based on this
investigation, we chose the region at 2.2070~$\mu$m $\pm$0.0050
(hereafter ``the Na interval'', see
Figure~\ref{fig-lambda.intervals}a) which includes two strong neutral
sodium lines (4s$^2$S$_{1/2}$--4p$^2$P$^0_{3/2}$), and several
prominent lines of neutral Si and Sc. We selected a second spectral
region at 2.2960~$\mu$m $\pm$0.0035 where the dominant feature is the
2--0 bandhead of $^{12}$CO (``the $^{12}$CO interval'', see
Figure~\ref{fig-lambda.intervals}b) for our analysis.
Figure~\ref{fig-na.behavior} presents a sequence of spectral syntheses
for the Na interval that illustrates the sensitivity of the line
ratios and line shapes in this wavelength band to the photospheric
temperature.

We produced a grid of spectra covering the appropriate range in
T$_{\rm eff}$ and $\log g$ using the NextGen non--grey atmosphere
models \citep{hauschildt1999}.  These models include TiO and H$_2$O
opacities, critical for cooler atmospheres.  We synthesized a high
resolution ($R=120,000$) K--band spectrum at 2.2 $\mu$m and 2.3~$\mu$m
using the MOOG spectral synthesis code \citep{sneden1973}.  Atomic and
CO line lists came from \citet{kurucz1994} and \citet{goorvitch1994},
respectively.  We have computed all models with solar metallicity.
The critical relative abundance in our analysis is [Sc/Si].  Stellar
abundances of [Si/Fe] in the local neighborhood are solar
\citep{edvardsson1993}, and we assume the same for [Sc/Fe].

For our analysis, we use synthesis models computed with solar
microturbulence values (1 km s$^{-1}$).  \citet{gray2001} have
compared spectral synthesis models to optical spectra of MK standards
deriving values of T$_{\rm eff}$, log~g, and microturbulence for stars
with spectral types from A5 to G2 and log~gs from 1.2 to 4.5.  At all
temperatures, they see a trend in the best--fit value of the
mircoturbulence. This value decreases as log~g increases, approaching
a roughly constant value at log~g~$>$~3.5.  This asymptotic value of
log~g decreases steadily from $\sim$2.5 km s$^{-1}$ at spectral type
A6 to $\sim$1.3 km s$^{-1}$ at spectral type G1, lending further
support to the appropriateness of our use of the solar microturbulence
value in our models of late--type stars with dwarf and sub--dwarf
gravities.

At cool temperatures (i.e. T$_{\rm eff}< 4000$ K), the wings of the
two Na~I lines are noticeably pressure broadened. The synthesis code,
MOOG, allows for the van der Waals damping parameter to be adjusted in
creating the artificial spectra.  We have tuned the amount of damping
present in the Na I lines to give the best fit to an observed spectrum
of the sun \citep{livingston1991}.  The best fit was the
\citet{unsold1955} approximation used to calculate the van der Waals
damping constant.

The strength of the Na lines increases with increasing surface gravity
at a fixed temperature, while CO lines decrease in strength. A
qualitative way to understand these trends is to examine how the
increase in electron presssure drives where the lines and continuum
form within the stellar atmosphere.  The greater electron pressures at
higher gravity results in a larger fractional abundance of neutral
sodium, causing the 2.2 $\mu$m lines to form closer to the stellar
surface.  This effect is larger than the decrease in the depth of
continuum formation with increasing electron pressure resulting in
larger line depths for Na.  In the case of CO, the increase in the
continuum opacity of H$^-$ is the dominant effect and the continuum
layer moves closer to the line forming region reducing the strength of
the bandhead absorption.
   
We also wish to derive the rotation rate ($v\sin i$), and the amount
of ``veiling'' (non--photospheric excess emission relative to the
photospheric flux).  At a given wavelength, for example that of the Na
interval, this veiling is defined as r$_{\rm Na}\equiv$~(F$_{\rm
source}-$F$_{\rm phot})/$F$_{\rm phot}$, where F$_{\rm source}$ and
F$_{\rm phot}$ are the observed flux and the photospheric flux,
respectively.  Therefore, we added extra dimensions to the T$_{\rm
eff}$ -- $\log g$ grid of synthetic spectra to include variations in
$v \sin i$ and r$_{\rm Na}$ as well.  To account for $v \sin i$
variations, we convolved the spectra with a rotational broadening
profile that had an assumed limb darkening coefficient of 0.6
\citep{gray1992} adding rotation at rates ranging from $v \sin i$ = 1
to 40 km s$^{-1}$.  The intensity of the non--photospheric emission
over the individual narrow spectral intervals has at most a slope of a
few percent, so we can treat it with a single parameter.  In the
presence of such an excess, resolved lines have lower equivalent
widths but retain the shapes imparted to them by line transfer in the
stellar atmosphere and by rotation. We alter the synthetic spectra to
include additional continuum to simulate veiling (r$_{\rm Na}$)
ranging from 0 to 8.

\subsection{Actual technique}

While, in all cases, our technique involves precise matching of
spectra synthesized from model atmospheres to high resolution
observations of limited portions of stellar spectra in the near--IR,
the exact procedure and which parameters we can derive depend on the
nature of the target stars and on the data available.  The interval
around the 2.2 $\mu$m sodium lines is particularly rich in diagnostic
power.  When observations of only this interval are available, we can
determine both the size of any excess emission and the rotation
velocity in PMS stars, as well as a quite well constrained measure of
the stellar effective temperature, all without recourse to stellar
photometry that is sensitive to reddening and to the circumstellar
excess.  We discuss our analysis for stars where only the Na interval
has been observed in $\S$~\ref{sec-na.alone}.

With the addition of observations of the $^{12}$CO interval, it is
possible to refine the T$_{\rm eff}$ determination and to determine
the surface gravity of the emitting star, even in the presence of
significant veiling and reddening.  In $\S$~\ref{sec-na.and.co}, we
explain how these improvements come about.  Throughout the discussion,
our focus is on applying the technique to reddened young TTSs
(DJW03). When using this method to characterize heavily reddened main
sequence stars, only minor modifications (such as dropping the free
parameter for near--IR veiling) are needed.
                                        
\subsubsection{Obtaining T$_{\rm eff}$, $v\sin i$, and r$_{\rm Na}$ from the 
  Na Interval Alone}\label{sec-na.alone}

We begin the spectral matching for the Na interval by using pattern
recognition to constrain the effective temperature to a 1000~K range.
For YSOs, we fix the value of $\log g =$3.5, which corresponds to
$\sim$1--2~Myr old objects in stellar evolutionary models
\citep{baraffe1998, siess2000, palla2000}, consistent with age
estimates of the central embedded cluster in Ophiuchus that includes
the sources studied in DJW03 \citep{wilking1989, greene1995,
luhman1999, bontemps2001}.  In $\S$~\ref{sec-errors} below, we
describe in detail how the choice of a particular value of $\log g$
affects the T$_{\rm eff}$ determination and how to correct T$_{\rm
eff}$ should $\log g$ have a different value.
 
Once we have the grid of spectral syntheses in place, we need to
choose spectral subintervals over which to compare the spectra
synthesized from the atmosphere models to the observed spectra.  Based
on our experience with both MK standards and YSOs, we restrict the
subintervals to the regions within the observed spectra where there is
measurable line absorption. The upper and lower boundaries of the
wavelength range vary with the apparent width of the stronger spectral
features.

\begin{figure}
\epsscale{0.8}
\plotone{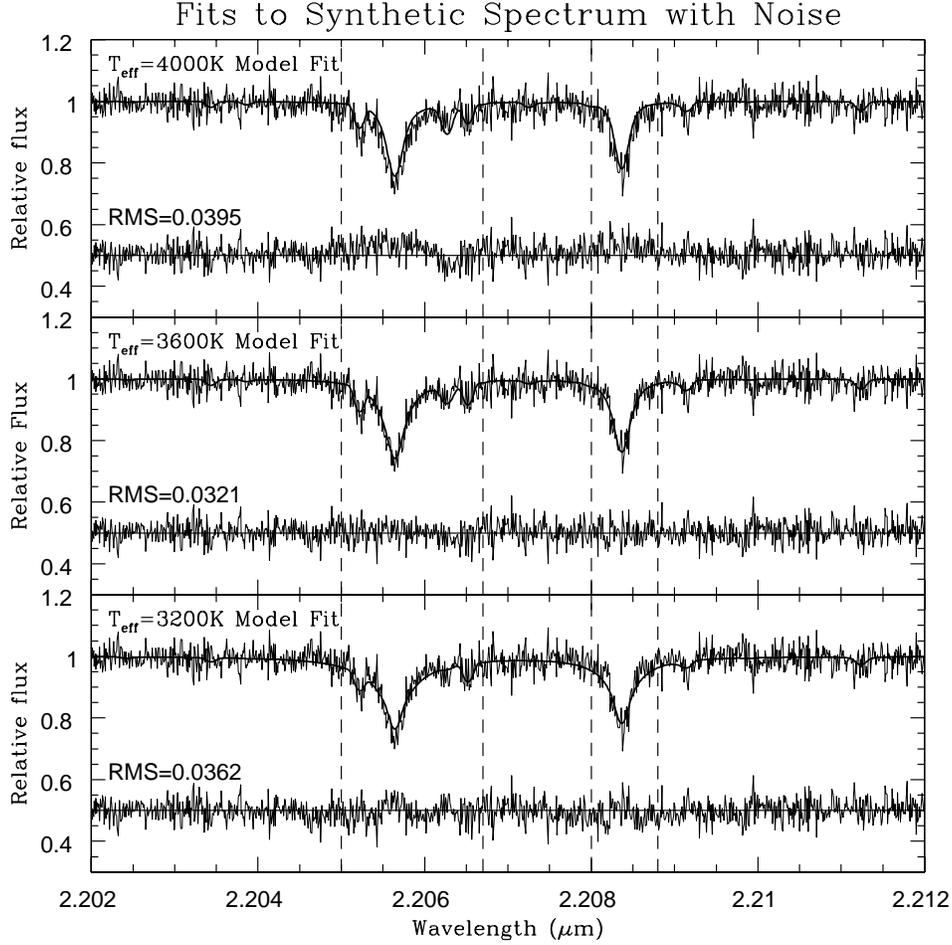}
\caption[Fits to a noisy artificial spectrum of the Na interval]
{\label{fig-fits.withnoise} Fits to a noisy artificial spectrum of the
Na interval.  The thin solid line shows an artificial spectrum for
T$_{\rm eff}= 3600$~K, $\log g= 4.0$, $v\sin i=15$~km~s$^{-1}$, and
r$_{\rm Na}=1.0$.  We have smoothed this spectrum to R~=~50,000 and
added Gaussian random noise to the spectrum to produce a S/N~=~30 (per
pixel with R~=~240,000 pixels).  Overlaid on each of the 3 panels is a
noiseless synthetic spectrum (bold lines) for $\log g= 4.0$, T$_{\rm
eff}=4000$K (top), T$_{\rm eff}=3600$K (middle), and T$_{\rm eff}=
3200$K (bottom). In each panel, the best fit was found holding T$_{\rm
eff}$ fixed at the listed value but allowing r$_{\rm Na}$ and $v\sin
i$ to vary.  The difference between the noisy artificial and noiseless
spectra is displayed at the bottom of each panel.  The vertical dashed
lines are the subintervals chosen in this case to enclose regions with
significant line flux over which we will compute the RMS difference
between the noisy artificial spectrum and the grid of noiseless
synthetic spectra. In the lower left of each panel is the RMS
difference between the noisy and noiseless spectrum normalized to the
number of points in the two subintervals that enclose the photospheric
lines.}
\end{figure}

The next step is to take the continuum normalized observed spectrum
and compare it to each synthesized model spectrum calculating the RMS
difference over the subintervals chosen in the previous step.
Figure~\ref{fig-fits.withnoise} illustrates the minimization by
showing an artificial noisy spectrum and how the differences between
this spectrum and the noiseless synthetic spectra vary as the search
routine steps through the correct value of T$_{\rm eff}$.  We then
find the combination of T$_{\rm eff}$, r$_{\rm Na}$, and $v \sin i$
values that would produce the minimum RMS difference between the model
and the observed spectra by interpolation.

\begin{figure}
  \epsscale{0.8} \plotone{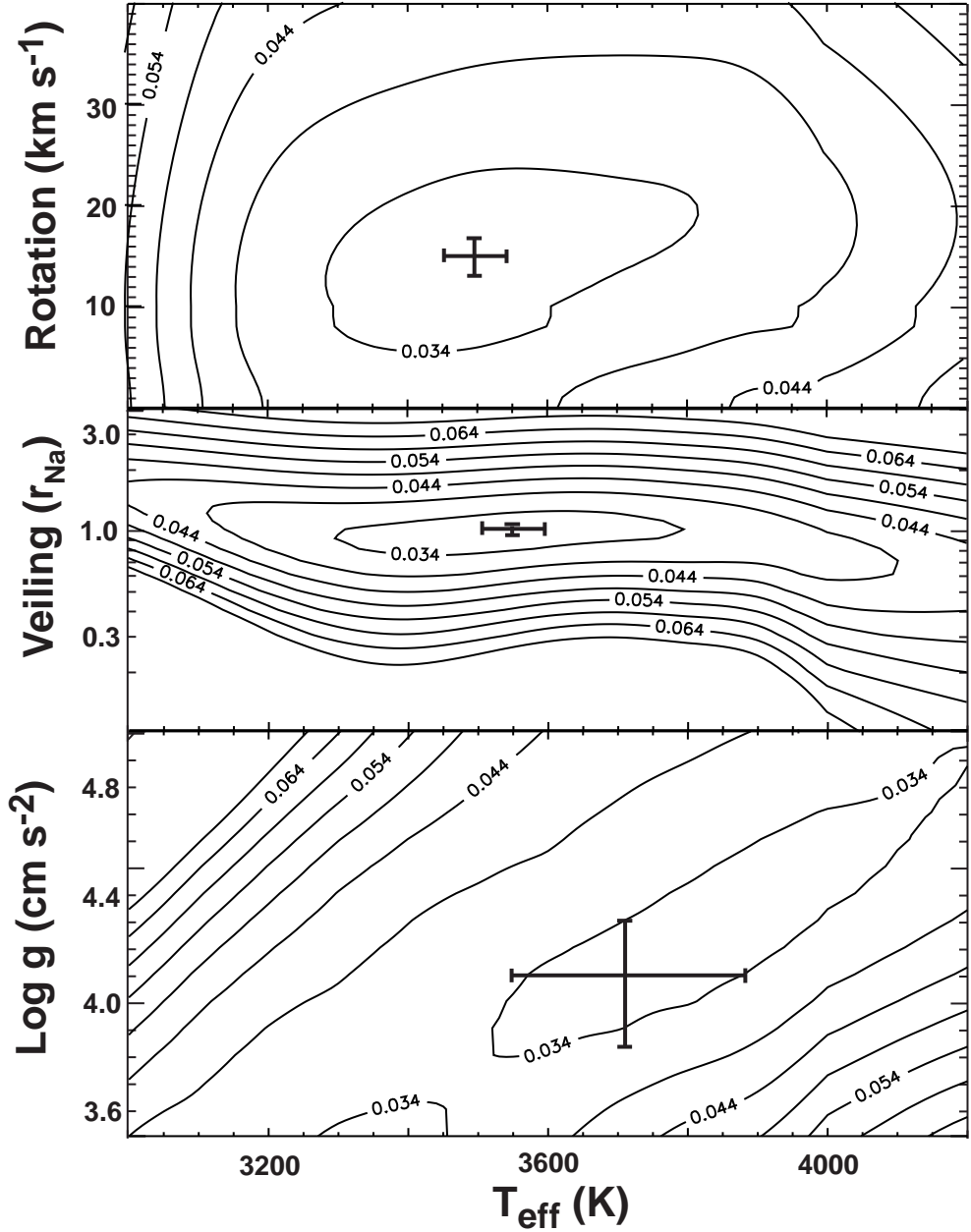}
\caption[Sections through an error hypersurface]
{\label{fig-contours} Variations of the RMS error in the T$_{\rm
eff}$--$v\sin i$, T$_{\rm eff}$--r$_{\rm Na}$, and T$_{\rm
eff}$--$\log g$ planes for a comparision of the noisy artificial
spectrum from Figure~\ref{fig-fits.withnoise} (where fluxes are
normalized to one) to noiseless synthetic spectra. The contours show
the RMS deviation of the noisy spectrum from the model at each point
in the parameter space.  We produced each two--parameter plot while
holding the other two variables fixed at the values matching the
correct values for the target spectrum.  The error bar at the RMS
minimum in each plot represents the standard deviation of the best fit
value for S/N~=~30 spectra with different noise seeds.}
\end{figure}

We illustrate the minimization process in Figure~\ref{fig-contours}
which shows the variation of the RMS difference between an artificial
noisy spectrum and various synthesis models of the Na interval as a
function of the search parameters.  We show variations of $\log g$
even though we would normally fix this parameter at a best--guess
value when only Na interval data are available.  The figure displays
the RMS difference and thereby the shape of the minimum in the T$_{\rm
eff}$--$v\sin i$ plane, the T$_{\rm eff}$--r$_{\rm Na}$ plane, and the
T$_{\rm eff}$--$\log g$ plane as the two variables are varied while
the other two variables are held fixed at their nominal values.  In
all three planes, there are well--determined minima in the fits to the
spectrum of the Na interval for all variables except $\log g$. The
minimum in the T$_{\rm eff}$--$\log g$ plane is very shallow.  It is
the one plane where we do not usually recover the input model.  The
grid in the $\log g$ direction is fairly coarse ($\Delta\log g=0.5$)
and some noise seeds for the artificial spectrum at $\log g=4.0$ even
find the lowest RMS value at the edge of the grid ($\log g=3.5$),
leaving the exact location of the minimum uncertain.  This cut (see
bottom panel of Figure~\ref{fig-contours}) illustrates that an
incorrect guess of $\log g$ for real target spectra can lead to
systematic errors in the derived value of T$_{\rm eff}$ (see
$\S$~\ref{sec-errors}).

\subsubsection{Solving Simultaneously for T$_{\rm eff}$ and $\log g$}\label{sec-na.and.co}
    
\begin{figure}
  \epsscale{0.8}
  \plotone{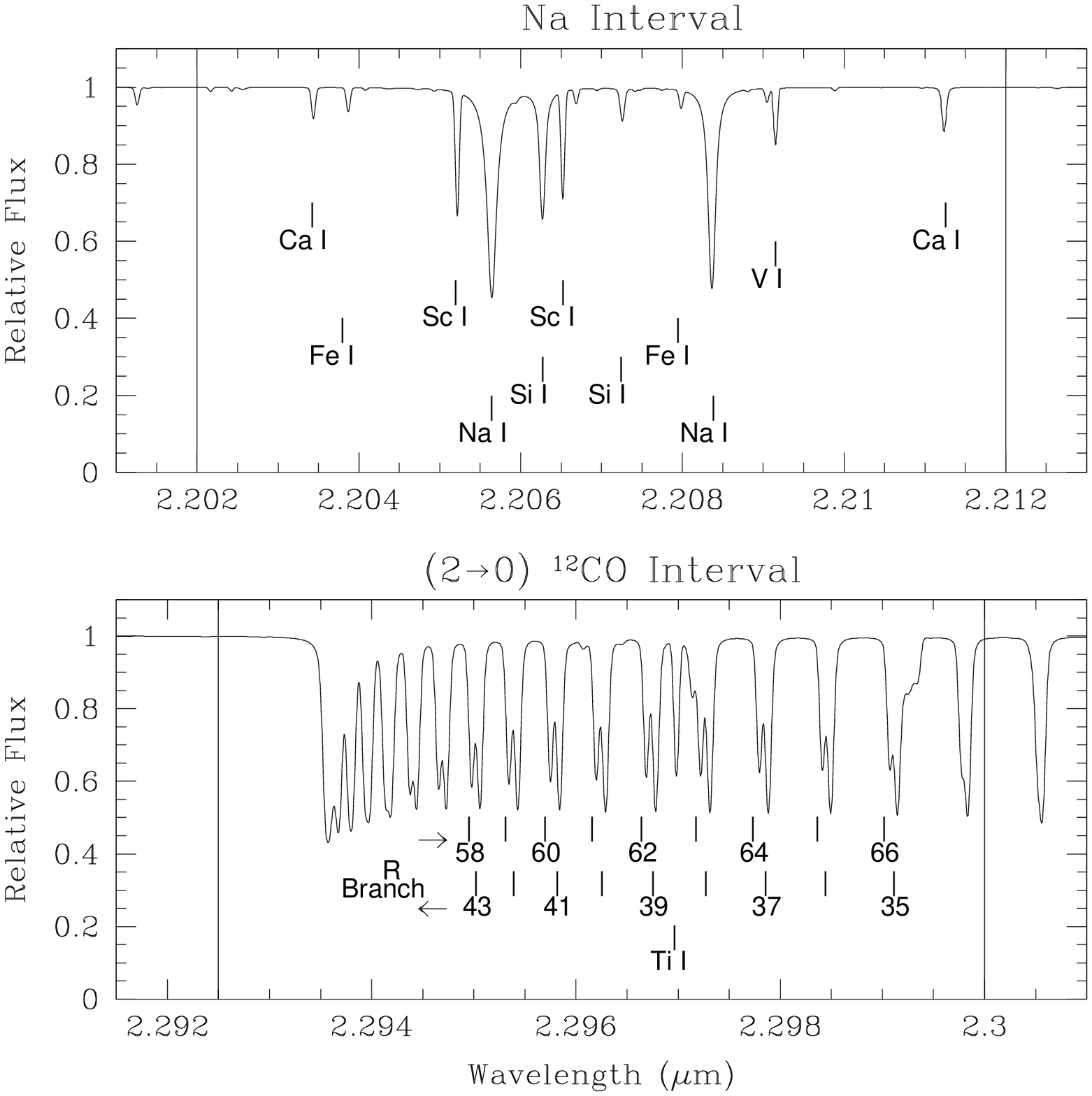}
\caption[Na and $^{12}$CO wavelength intervals]
{\label{fig-lambda.intervals} The intervals within the 2.0--2.4~$\mu$m
K window used for spectral synthesis analysis of high resolution
data. The spectra are from a synthesis of a T$_{\rm eff}=4000$~K,
$\log g=3.5$ NextGen atmosphere model \citep{hauschildt1999}, assuming
a resolving power $\lambda$/$\Delta \lambda = 50,000$. The intensity
is normalized to the photospheric continuum.  The Na interval shows
the numerous neutral atomic species present in the photospheres of
cool stars (top). The $^{12}$CO interval shows the bandhead and the
v=2--0 R--branch transitions of $^{12}$CO (bottom).  The vertical
lines show the boundaries for the intervals used to compute the
equivalent width ratios plotted in Figure~\ref{fig-isograv}.}
\end{figure}

There is a strong inverse dependence of the (2--0)~$^{12}$CO bandhead
equivalent width on $\log g$.  There is also a weaker but noticeable
dependence of the line equivalent width for the Na interval with $\log
g$.  Figure~\ref{fig-isograv} plots the ratio of (2--0)~$^{12}$CO to
Na interval photospheric equivalent width (the equivalent width summed
over the lines within the interval after removal of any
non--photospheric continuum from the spectrum (see Appendix A) as a
function of T$_{\rm eff}$ for $\log g$ values ranging from 3.5 to 5.0.
We derived this ratio from the NextGen photospheric models by creating
synthetic spectra for the relevant intervals and integrating the
spectra over the bands marked in Figure~\ref{fig-lambda.intervals}.
For T$_{\rm eff}>$ 3700~K, the ratio varies strongly with $\log g$ and
is almost independent of temperature.  For lower temperatures, the
ratio still varies strongly with $\log g$ but T$_{\rm eff}$ must also
be known to correct for the sensitivity of the equivalent width ratio
to temperature.

\begin{figure}
  \epsscale{0.8}
  \plotone{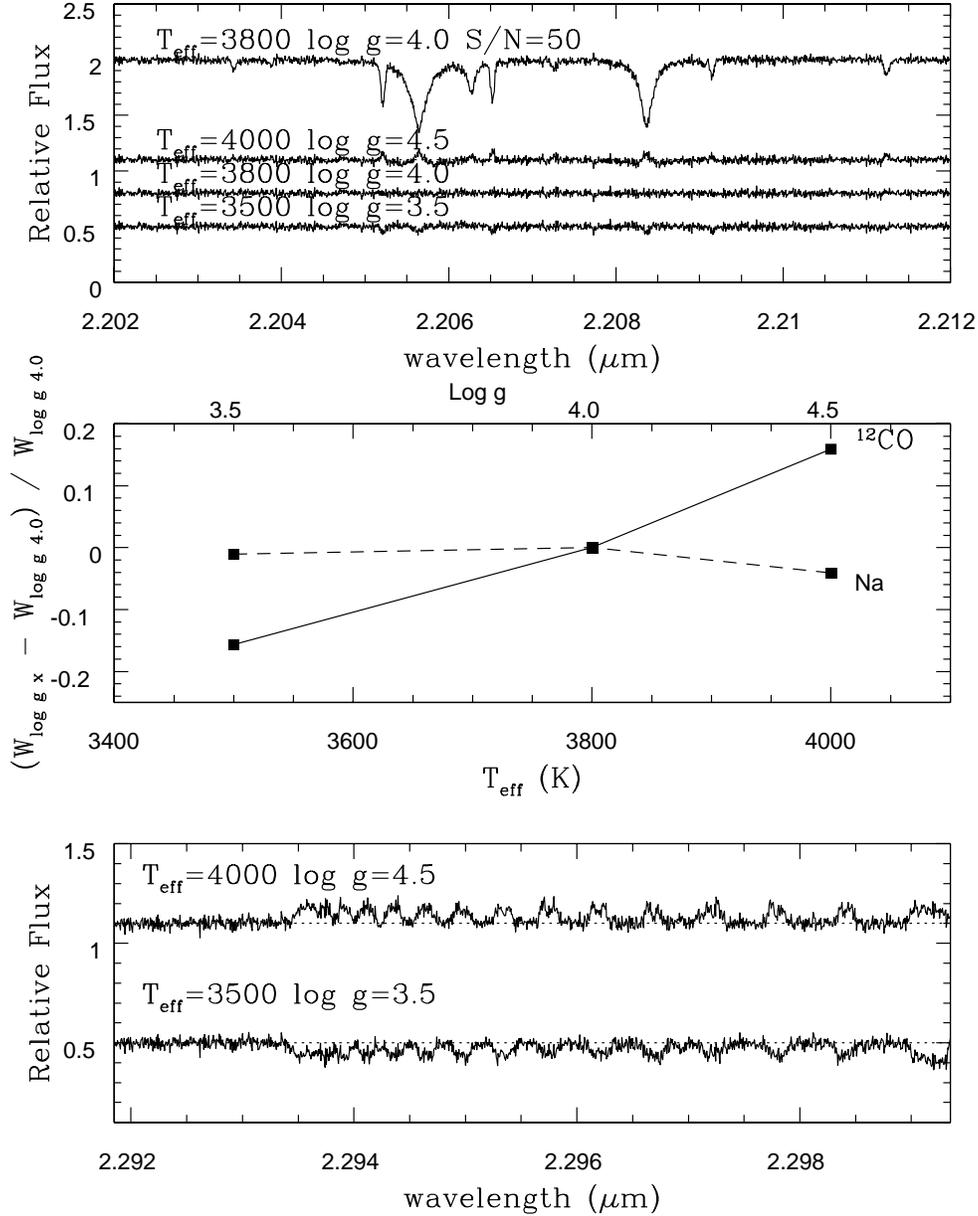}
\caption[Demonstration of simultaneous derivation of T$_{\rm eff}$ and
$\log g$]
{\label{fig-loggteff.degen} Demonstration of simultaneous derivation
of T$_{\rm eff}$ and $\log g$.  The top panel shows a S/N~=~50
synthetic spectrum for (T$_{\rm eff}$, $\log g$)= (3800K, 4.0).  Below
that we show differences between this spectrum and models for (T$_{\rm
eff}, \log g)=$~(3500K, 3.5), (3800K, 4.0), and (4000K, 4.5)
illustrating the effects of the broad minimum in the errors as T$_{\rm
eff}$ and $\log g$ increase simultaneously.  The middle panel shows
the difference between the equivalent width of the Na and CO interval
for these (T$_{\rm eff}, \log g)$ pairs and (T$_{\rm eff}, \log
g)=$~(3800K, 4.0), normalized to the equivalent width of the (3800K,
4.0) spectrum.  The bottom panel shows the associated strong
variations in the spectra of the CO interval.}
\end{figure}

In $\S$~\ref{sec-na.alone}, we derived T$_{\rm eff}$, $v\sin i$, and
r$_{\rm Na}$ from spectra of the Na interval while holding $\log g$
fixed.  When we have data available for both the Na and the $^{12}$CO
intervals, we are able to solve for $\log g$ rather than just assume
its value. We determine $\log g$ and the other parameters iteratively.
We begin by assuming $\log g=3.5$ and derive a best fit for T$_{\rm
eff}$, r$_{\rm Na}$, and $v\sin i$ using the Na observations as in
$\S$~\ref{sec-na.alone}.  We take the values of T$_{\rm eff}$ and
r$_{\rm Na}$ that this process produces and use them, together with
the integrals over the spectral intervals shown in Figure
\ref{fig-lambda.intervals}, to compute the ratio of photospheric
equivalent widths in the $^{12}$CO and Na intervals (see Appendix~A).
With the photospheric equivalent width ratio and the derived value of
T$_{\rm eff}$, we can use the relations plotted in
Figure~\ref{fig-isograv} to estimate $\log g$ and use this new value
of $\log g$ in an iteration of the procedure for deriving T$_{\rm
eff}$, $v\sin i$ and r$_{\rm Na}$ from the observed spectrum of the Na
interval.  The iterative procedure converges quickly on a value for
$\log g$.  The first two panels of Figure~\ref{fig-loggteff.degen}
illustrate how the derivation of parameters from the Na interval plus
the use of the $^{12}$CO interval photospheric equivalent width work
together to produce the correct value for all four parameters.  The
figure shows how the equivalent width of the CO bandhead varies
significantly more than the equivalent width of the Na interval going
from (T$_{\rm eff}$, $\log g$)~=~(3600 K, 3.5) to (4000 K, 4.5). Note
that at low temperatures and high surface gravities, the Na line wings
extend beyond the spectral interval over which we compute the
equivalent width for the Na interval.  When the data cover a similar
or narrower spectral interval, observations of cool and/or high
surface gravity stars must be corrected for the fact that the
intensity at the edges of the band is not fully at the level of the
photospheric continuum. We have applied this correction in
Figure~\ref{fig-isograv}.  Therefore even when data with broader
spectral coverage are available, the measured equivalent width should
be computed only over the marked intervals in
Figure~\ref{fig-lambda.intervals} for comparison to
Figure~\ref{fig-isograv}.

Although our procedure for using the CO interval together with the Na
interval to determine $\log g$ does not, in principle, require high
spectral resolution observations of the CO bandhead, such observations
are very useful.  In the youngest YSOs, the excess non--photospheric
emission can be many times greater than the emission from the
photosphere itself.  In such cases, the equivalent widths of the lines
not only fail to represent the conditions in the stellar atmosphere
but also can be extremely hard to measure.  In low resolution spectra,
the combination of dilution by non--photospheric emission and dilution
because the features are unresolved can make reliable measures of
equivalent widths very problematic.  Good atmospheric cancellation is
also difficult because of the inherent messiness both of the stellar
and the telluric spectrum in the region of the (2--0)~$^{12}$CO
bandhead.  Higher resolution spectra improve the situation because
they make it easier to cancel telluric lines and because
line--to--continuum ratios are larger.

At high resolution, where we can resolve the CO bandhead and also the
adjacent features in the ascending and descending R--branch and where
effective telluric line cancellation is possible, we can match the
depth and shape of a CO bandhead to obtain useful information about
both T$_{\rm eff}$ and $\log g$ that does not depend strongly on
precise determination of the continuum level.

The third panel in Figure~\ref{fig-loggteff.degen} illustrates the
dependence of the spectral shape in the (2--0)~$^{12}$CO bandhead on
$\log g$ and T$_{\rm eff}$ as it would be seen at high spectral
resolution.  This panel demonstrates the potential for using detailed
spectral shapes in the $^{12}$CO interval to improve the robustness of
the stellar parameter derivation scheme.  It shows the difference
between the synthetic spectrum for the CO interval for (T$_{\rm eff}$,
$\log g$)=(3800~K, 4.0) and (4000~K, 4.5) and (3600~K, 3.5).  We see
that, in addition to the equivalent width changes, there is a change
in the individual line depths and in the shape of the envelope of the
bandhead at this resolving power (R~$=50,000$).  In the current work,
however, we restrict ourselves to analyzing the results of detailed
spectral synthesis matching for the Na interval combined with use of
the Na/$^{12}$CO photospheric equivalent width ratio to arrive at
accurate values for T$_{\rm eff}$, $v\sin i$, r$_{\rm Na}$ and $\log
g$.

\section{Analysis of Errors}\label{sec-errors}

Now that we have developed a procedure for determining stellar
parameters from high resolution spectra, we would like to know how
well it works, both the sensitivity of the fitting routine and the
uniqueness of the derived solutions.  We discuss here four classes of
uncertainty: (1) The sensitivity of the model fitting to random noise
in the spectra. (2) Internal systematic uncertainties arising from the
optimization scheme and from degeneracies between various stellar
parameters. (3) External systematic errors introduced by
transformations from one theoretical framework to another. (4) Errors
arising from non--random effects present in the data.  We evaluate the
effects of these problems on derived parameters using simulations,
real data, and modifications of real data.  We first analyze the
errors in parameters derived using only the Na interval and then
discuss changes in these errors and the uncertainty in the
determination of $\log g$ when observations of both the Na and
$^{12}$CO intervals are available.

\subsection{Random Errors}

\begin{figure}
  \epsscale{0.8}
  \plotone{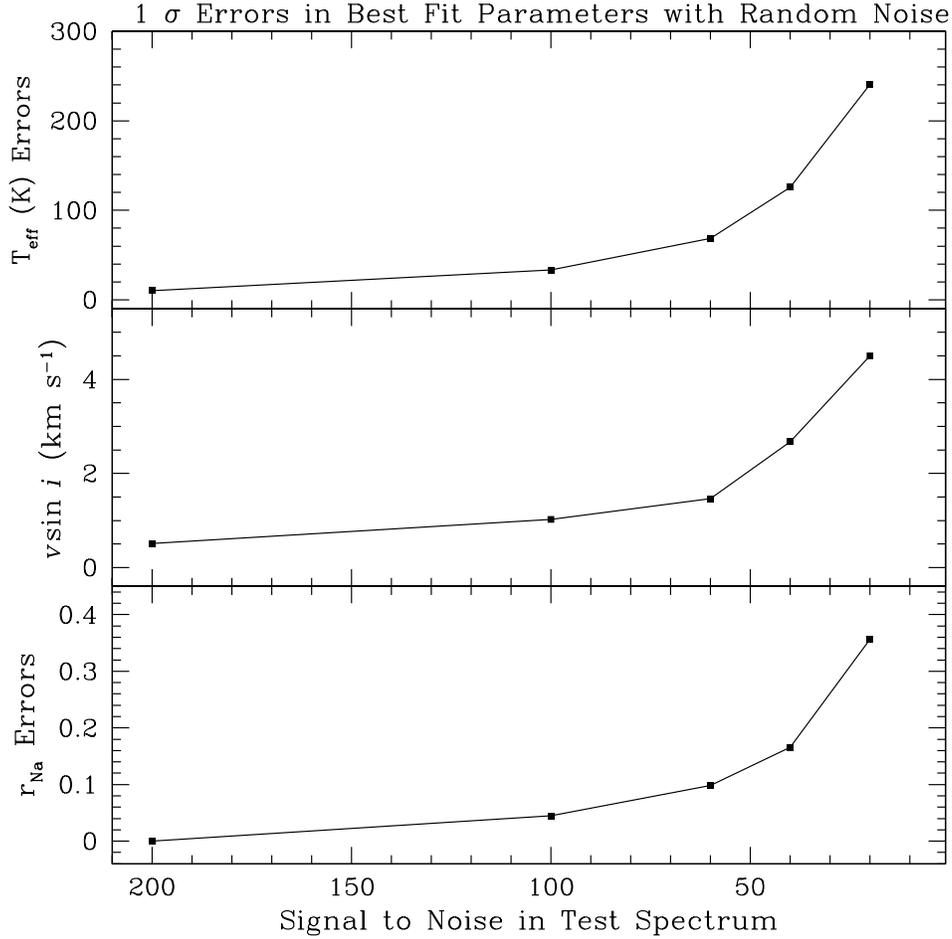}
\caption[Standard deviation of the derived value of T$_{\rm eff}$ as a
function of S/N] 
{\label{fig-noise.tefferrors} Standard deviation of the derived value
of T$_{\rm eff}$ (derived using the Na interval only) as a function of
S/N.  To construct this plot, we used a noisy artificial spectrum with
T$_{\rm eff}$=4200K, $\log g$=4.0, $v\sin i$=15 kms$^{-1}$, and
r$_{\rm Na}$=1.0 as the target and fit for the best temperature
allowing all parameters except $\log g$ to vary.  We compared the
target and model spectra at sampling points spaced every $\Delta
\lambda$/$\lambda$=(1/240,000) along spectra smoothed to a resolving
power of 50,000.  The S/N is that appropriate to data binned to
channels $\Delta \lambda$/$\lambda$=(1/50,000) wide.}
\end{figure}

A generalized assessment of the sensitivity of our fitting routines to
random errors is not possible.  The sensitivity of the T$_{\rm eff}$,
$v\sin i$, and r$_{\rm Na}$ determinations will be not only a function
of the signal--to--noise ratio (S/N) of the spectra but also will
depend on the widths of the lines and on their strengths relative to
the continuum (that is, effectively on all four parameters). By
working with a typical spectrum, however, we can get a rough idea of
what S/N we require to reach the point where random noise in the
spectra no longer dominates the uncertainty in determining stellar
parameters.  We illustrate the sensitivity to random errors by taking
a synthetic spectrum for the model shown in
Figure~\ref{fig-lambda.intervals}a and adding more and more Gaussian
random noise to it.  We take each artificially noisy spectrum, find
the best--fit values for T$_{\rm eff}$, $v\sin i$ and r$_{\rm Na}$ in
the usual way, by calculating the RMS difference between this spectrum
and a set of noise--free synthesis models covering a range in all
three parameters and then interpolating to obtain the best--fit
values.  At each S/N level, we re--seed the noisy spectrum 30 times
and repeat the fitting procedure.  The standard deviation about the
mean value of each derived parameter for this ensemble of noisy
spectra then reflects the uncertainty due to random noise at a given
S/N level.  Figure~\ref{fig-noise.tefferrors} shows how the standard
deviation about the mean derived T$_{\rm eff}$, $v\sin i$, and r$_{\rm
Na}$ varies for noisy versions of the Na interval spectrum in
Figure~\ref{fig-lambda.intervals}a (with continuum added to make
r$_{\rm Na}=1$) as the S/N decreases.  For this particular case, the
uncertainty in the T$_{\rm eff}$ due to random errors is less than
100~K for S/N $>$ 50.  Random errors result in uncertainties of less
than one spectral subclass as soon as the S/N ratio is greater than 35
at R~$=50,000$.  At S/N~=~50, the random uncertainty in $v\sin i$ is 2
km s$^{-1}$ and that in r$_{\rm Na}$ is 0.13.  The curves shown in
Figure~\ref{fig-noise.tefferrors} illustrate the behavior of the
random errors with S/N for a spectrum with r$_{\rm Na}$=1.  For
sources with other values of r$_{\rm Na}$, we can derive the S/N
required for a given uncertainty in T$_{\rm eff}$, $v\sin i$, or
(1+r$_{\rm Na}$) by multiplying the value shown in
Figure~\ref{fig-noise.tefferrors} by (1+r$_{\rm Na}$)/2.  For a given
uncertainty in (1+r$_{\rm Na}$), the uncertainty in r$_{\rm Na}$
itself is (1+r$_{\rm Na}$) times larger.  Once the S/N exceeds
$\sim$35$\times$(1+r$_{\rm Na}$), other forms of errors begin to
dominate the uncertainty in the derivation of stellar parameters from
the K--band spectral fitting technique.

For $v\sin i$, we also tested the dependence of the uncertainty on
T$_{\rm eff}$.  We used models with stellar rotation (25~km~s$^{-1}$),
instrumental smoothing (R $=50,000$) and Gaussian noise (S/N $=30$ per
R $=240,000$ channel with r$_{\rm Na}$=0) added to simulate real data.
We fit the modified synthetic spectra to a set of noiseless models
keeping the temperature fixed at the correct value and allowing the
RMS minimization algorithm to select the best $v\sin i$ value.  This
was repeated with 30 noise seeds for each temperature giving a mean
and 1 $\sigma$ error for $v\sin i$ for temperatures between 3200 K and
4400 K.  The 1$\sigma$ error is less than 2 km s$^{-1}$ over the
entire temperature range.

\subsection{Uncertainties Arising from Internal Systematics}

For some pairs of stellar parameters, changes of both parameters
simultaneously in a certain sense can keep the line shapes and depths
almost unchanged. These partial degeneracies in the output line shapes
between different groupings of stellar parameters can have the effect
of exaggerating the uncertainties caused by random errors.  For
minimizations of model--observed spectral differences in the Na
interval, this effect is most evident in the broadness of the minimum
RMS error along a diagonal in the T$_{\rm eff}-\log g$ plane
(Figure~\ref{fig-contours}, bottom).  When the spectra are noisy or
imperfect, there is a range of temperature--gravity pairs with very
similar RMS values.  Similar degeneracies for other lines in the
near--IR have caused difficulties when attempting to type YSOs from
low resolution near--IR spectra, unless one knows what luminosity
class is appropriate for the template stars \citep{luhman1999}.

\begin{figure}
  \epsscale{0.8}
  \plotone{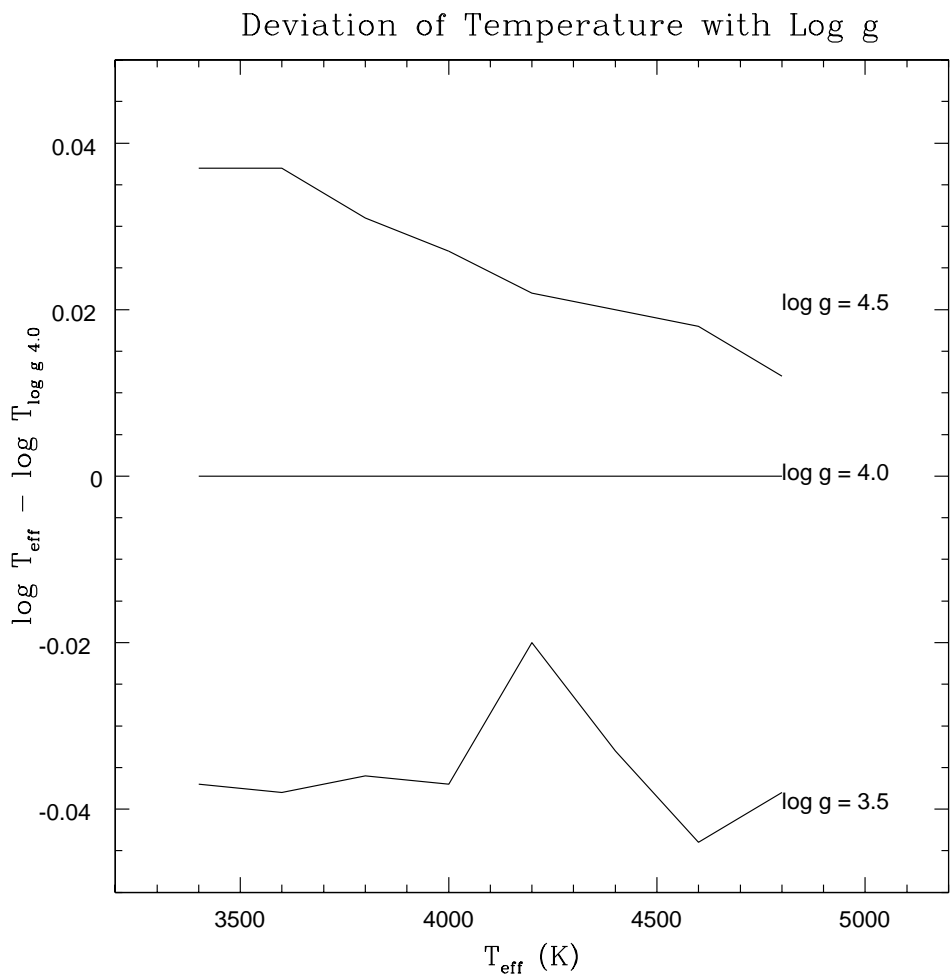}
\caption[The effect on the best--fit value of T$_{\rm eff}$ of
assuming an incorrect value for $\log g$] 
{\label{fig-logg.degenfix} The effect on the best--fit value of
T$_{\rm eff}$ of assuming an incorrect value for $\log g$.  At each
temperature, a synthetic target spectrum with $\log g$=4.0 and two
grids of MOOG models with $\log g$=3.5 and $\log g$=4.5 were created.
We then fit for the temperature of the target spectrum using each of
the grids with the different values of $\log g$.  The solid lines show
the difference between the logarithm of the best--fit T$_{\rm eff}$ to
the target spectrum derived by using the $\log g$=3.5 (bottom) or 4.5
(top) grids and the logarithm of the target spectrum T$_{\rm eff}$.}
\end{figure}

The similar line shapes for models along a diagonal in the
temperature--gravity plane mean that we can introduce systematic
errors in the derived T$_{\rm eff}$ when we assume a value of $\log g$
and then derive the temperature.  We can illustrate and assess this
effect with fits to models.  Figure~\ref{fig-logg.degenfix} summarizes
the results of our tests by showing how the derived temperatures
deviate from the target spectrum temperature at different target
values of T$_{\rm eff}$. Typically, our best fits for T$_{\rm eff}$
using models with a $\log g$ differing by $\pm 0.5$ from the $\log g$
of the target spectra mis--estimate the temperature by about 7\%.
This difference is approximately 1--2 spectral subclasses over the
range of T$_{\rm eff}$ relevant to our study. If we later obtain an
independent estimate of $\log g$, we can correct the derived T$_{\rm
eff}$.  For data where we derive an effective temperature from the Na
interval assuming some value of $\log g$, we can correct the derived
$\log$~T$_{\rm eff}$ by $+$0.06 for every $\log g=1$ difference
between the assumed and the correct values.  Making this correction,
we recover the actual value of T$_{\rm eff}$ to within better than
4\%.

The way in which we create the error space from comparisons of
observed target spectra and synthesized model spectra may have a
systematic effect on the derived parameters.  The error space shown in
Figure~\ref{fig-contours} represents the RMS deviation of the target
spectrum from the models over selected intervals where line absorption
was present (see Figure~\ref{fig-fits.withnoise}).  This scheme allows
for variations of feature depths and shapes.  Because it is a straight
RMS, it weights the stronger features, in particular the Na features,
more heavily.  One might ask if this is the best scheme, i.e. does it
make the uncertainties in the derived parameters larger than they need
to be?  In its favor is the ability of the Na line depths to
distinguish the value of r$_{\rm Na}$ and the sensitivity of the depth
and shape of these lines to T$_{\rm eff}$ and $\log g$.  On the other
side of the ledger is the low weight a straight RMS gives to the
weaker Sc and Si lines.  The ratio of these lines is the most
sensitive temperature indicator in the Na interval.  More complex
schemes that make better use of the information content of the weaker
lines and the subtleties of the line shapes are certainly possible.
When the S/N gets large enough that systematic errors dominate over
random errors, a straight RMS does not do as well as a weighted error
scheme, but remains a reasonable approximation.  The simple RMS
scheme, however, has the great advantage that it finds the right
parameters robustly over our whole temperature range at various values
of r$_{\rm Na}$ and $v\sin i$ and that its level of complexity is
appropriate to S/N ratios of 30--50 where random errors are just
beginning to give way to systematics.

\subsection{Errors in the Radial Velocity Determination}

When we work with real data, we use the RMS minimization of the
difference between the data and the synthesis models to determine the
best radial velocity shift of the observed stars along with the
best--fit stellar parameters.  A narrow search range in radial
velocity space is first selected by inspection.  The data and models
are interpolated to a higher dispersion (R$_{\rm pix}=360,000$). We
then use the minimum RMS of the residuals to the fit to select the
best sub--pixel radial velocity shift.  Tests with artificial spectra
show that random errors in the radial velocity determination due to
noise in the spectrum are small ($< 0.5$ km s$^{-1}$) for spectra with
S/N $>$ 30 per pixel (R$_{\rm pix}=240,000$).

We also examined how errors in the radial velocity fit affect our
determination of the stellar parameters. For a S/N $\sim$30 per pixel,
noise in the spectra is a much more significant factor in causing
errors in $v\sin i$ than the error in the radial velocity
determination.  Stellar parameters determined mostly from line depths
(i.e. effective temperature and veiling), are not sensitive to radial
velocity errors at the 5 km s$^{-1}$ level.

\section{Comparison with Standards and External Systematics}

We can use the spectra we have taken in the 2.2 $\mu$m Na interval of
MK standard stars to perform a real--world test of our fitting
technique.  We discuss here the test results for T$_{\rm eff}$ and
$v\sin i$.  We also add an artificial infrared excess and rotation
velocity to the MK standards in order to look for systematic effects
in the determination of r$_{\rm Na}$ and refine our understanding of
such effects on our determination of $v\sin i$.

The high resolution spectra we use to derive stellar parameters are
imperfect representations of the source spectra.  Systematic problems
with the data include imperfect flat--fielding, defects in the
cancellation of telluric features and the presence of scattered light
or leaked out--of--order stellar radiation.  The use of high
resolution spectra in our analysis reduces the effect of these
problems substantially.  By choosing restricted wavelength intervals
over which to match the synthetic models to the data, we minimize
flat--fielding effects since mismatches in the shapes of the synthetic
and observed lines then play a bigger role in influencing the RMS
difference.  At high spectral resolution, the residuals of telluric
lines cover a limited and known part of wavelength space.  We simply
exclude these regions from our fitting spectral matching intervals.
Scattered light is usually removed in the data analysis if the source
and sky have been switched regularly between two positions along the
slit.

The continuum level for our spectra could be another free parameter in
our spectral fitting routine, though we have chosen to leave the
continuum fixed when finding the minimum RMS difference between the
target and synthetic spectra.  For our target spectra, we have
normalized the continuum based on a linear fit of two points close to
the edges of the spectrum and farthest from Na I lines where the
potential influence by any damping wings is minimized.  In some cases
where the instantaneous spectral coverage of the spectrometer is
limited, it can be difficult to set the continuum level in the spectra
correctly, particularly for the cooler stars and stars with higher
surface gravities where the wings of the Na lines are extended.

Differences between the effective temperatures derived from optical
spectral types and the T$_{\rm eff}$ values we derive from high
resolution observations of the Na interval reflect the sum of the
internal errors in our determination of T$_{\rm eff}$, errors in
spectral typing from optical observations, and errors in converting
from spectral types to effective temperatures. This last error can be
substantial.  De Jager \& Nieuwenhuijzen (1987) estimate errors of
0.021 in the $\log $ of T$_{\rm eff}$ in their spectral type--T$_{\rm
eff}$ conversion for dwarfs, corresponding to $\pm$200~K at T$_{\rm
eff}$=4000~K.

\begin{figure}
  \epsscale{0.8} \plotone{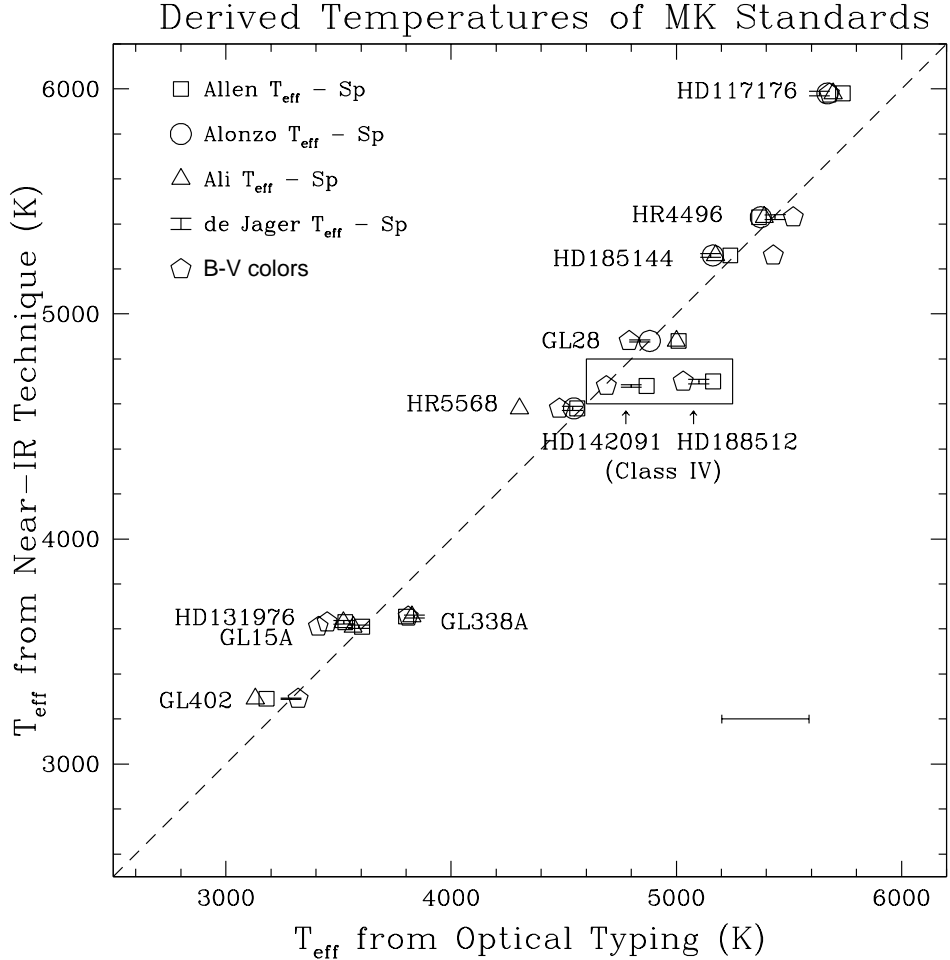}
\caption[Test of our effective temperature determinations against
temperature determinations available in the literature for MK
standards] 
{\label{fig-mk.stands} Test of our effective temperature
determinations (vertical axis) against temperature determinations
available in the literature for MK standards (horizontal axis). The
small vertical error bars represent $\pm$1$\sigma$ uncertainties due
to the noise in the observed spectra, determined by calculations
similar to those used in Figure~\ref{fig-noise.tefferrors} at each
temperature and at a S/N comparable to that of the data.  The
horizontal error bar in the lower right corner of the Figure shows the
1 $\sigma$ uncertainty in the spectral type to T$_{\rm eff}$
conversion at 4000~K \citep{dejager1987}.  We placed the symbols in
the x direction by converting the spectral types
(Table~\ref{tbl-mk.stands}) to T$_{\rm eff}$ using various spectral
type--T$_{\rm eff}$ relations: open circles \citep{alonso1996}, open
triangles \citep{ali1995}, open squares \citep{allen2000} vertical
error bars \citep{dejager1987}.  Unless otherwise marked, all sources
are dwarfs, luminosity class V.  The open pentagons show the T$_{\rm
eff}$ derived from the reported B--V colors using the conversion
relation of \citet{kenyon1995}. The dashed line shows the relation
T$_{\rm eff}$(near--IR) = T$_{\rm eff}$(optical).}
\end{figure}

Table 1 lists the effective temperature we derive for MK standards in
our sample using the Na interval, as well as temperatures derived from
the optical spectral types using the T$_{\rm eff}$--spectral type
relation of \citet{dejager1987}. In Figure~\ref{fig-mk.stands}, we
plot the near--IR T$_{\rm eff}$ against T$_{\rm eff}$ determined from
optical measurements by a variety of different techniques.  For a
given target, the spread of different symbols along the horizontal
axis illustrates differences in the conversion between spectral type
and temperature \citep{dejager1987, ali1995, alonso1996, allen2000}.
The figure also includes temperatures derived using observed B--V
colors and the conversion relation of \citet{kenyon1995} which we will
also apply to YSOs.  The vertical error bars (placed on the de Jager
T$_{\rm eff}$ points) show our estimate of the 1$\sigma$ uncertainty
due to random noise in our data.  The errors for our T$_{\rm eff}$
determinations were derived by carrying out an analysis similar to
that used to create Figure~\ref{fig-noise.tefferrors} but at the
temperature and S/N ratio appropriate to each MK standard.  The size
of these errors indicates that systematic effects must dominate any
differences between the optical and infrared results.

For the luminosity class V MK standards, a best fit to the T$_{\rm
eff}$ \citep[optical,][]{dejager1987} versus T$_{\rm eff}$(Na
interval) yields:

\begin{equation}
\label{eqn-optir.Teff}
{\rm T_{eff}(Na)}= 1.08 \times {\rm T_{eff}}(\rm{de~Jager}) - 295
\end{equation}

The 1$\sigma$ deviation from this relation is 113 K.  For the same
sources, the 1$\sigma$ deviation for an ``assumed'' relation of
T$_{\rm eff}$(Na)=T$_{\rm eff}$(de Jager) is 141 K, which is still
less than the quoted uncertainty in the spectral type--temperature
conversion.

All of the MK standards in our sample are fairly slow rotators and
therefore not particularly useful for tests of the accuracy of our
fits to the Na interval for $v\sin i$.  Table 1 lists the values of
$v\sin i$ from optical spectra as well as the results of our fit to
the Na interval. For these near--IR results, we have removed the
effect of slit broadening by subtracting it in quadrature from the
fitted value to produce the intrinsic widths or upper limits listed in
the Table.  For 10 of the 11 stars, the optical and near--IR
measurements and upper limits are consistent with each other. For the
one discrepant source, HD~131976, where we measure $v\sin i$= 10 km
s$^{-1}$, versus the value of 1.4 km s$^{-1}$ measured by
\citet{duquennoy1988}, the S/N for the infrared measurement is lower
than that of any other standard in our sample.

When we study a particular embedded PMS star, we usually do not know
its metallicity.  Assuming solar metallicity, however, should not
cause problems for comparisons between models and stellar spectra
since the deviations from solar metallicity are usually quite small,
typically $\Delta$[Fe/H]~$<$~0.1 \citep{padgett1996}.  For the MK
standards in the field, differences between the assumed and actual
stellar abundance can cause systematic differences in the T$_{\rm
eff}$ scale.  Eight of the MK standards we have observed as a test
sample have measured abundances ranging from +0.02 dex above solar to
$-$0.30 dex below (Table~\ref{tbl-mk.stands}).  For the comparison
with MK standards in Figure~\ref{fig-mk.stands}, we fixed r$_{\rm
Na}$=0 and used the metallicities listed in Table~\ref{tbl-mk.stands}
or solar metallicities when no measurements were available.  For MK
standards with measured non--solar metallicities, we computed separate
grids of synthetic spectra for comparison with the observed spectra.
Since the stellar atmosphere models available to us were gridded
rather coarsely in metallicity for our purposes (every 0.5 dex), we
constructed the spectral synthesis grids for mildly metal--poor stars
($0.3 \leq [{\rm Fe/H}] \leq 0$) using solar metallicity model
atmospheres but synthesizing the spectra using abundances scaled down
by [Fe/H].  This procedure does not account perfectly for metallicity
effects.  If we constrain the search grid to a fixed veiling (r$_{\rm
Na}=0$) and metallicity determined from the literature, however, the
best fit model (as determined by the minimum RMS difference) has a
noticeably non--zero residual equivalent width when compared to the
observed spectrum of the MK standard.

In order to test the ability of our fitting routine to derive $v\sin
i$ and r$_{\rm Na}$ as well as T$_{\rm eff}$ from data containing
realistic amounts of systematic deviation from ideal spectra, we have
altered our PHOENIX observations of MK standards
($\S$~\ref{sec-observations}).  We began with MK standards with very
small rotation velocities ($v\sin i \leq 10$ km s$^{-1}$).  These
stars also had no intrinsic veiling (r$_{\rm Na}$=0). As we showed
above, if we hold r$_{\rm Na}$ fixed at zero, we recover the T$_{\rm
eff}$ derived from optical spectra from our near--IR observations. To
each spectrum, we then added a known amount of rotation ($v\sin i$ =
25 km s$^{-1}$) and veiling (tripling the amount of continuum to make
r$_{\rm Na}$ = 2.0).  The Na interval spectra of these doctored stars
were then analyzed using our standard procedure. For 6 of the 7
objects, we recovered effective temperatures close to those derived
from the unaltered stellar spectra, typically within one spectral
subclass of the best--fit temperature for the unaltered spectrum
(Table~\ref{tbl-mkstands.veiling}).  The recovered temperature tended,
however, to be systematically lower than the values derived holding
r$_{\rm Na}$ fixed and the recovered r$_{\rm Na}$ values were higher
than those we put into the spectra.  The seventh object, HD~117176,
has a T$_{\rm eff}$ higher than the range for which the technique is
fully reliable.  For this source, the fitting routine derived a lower
temperature and higher veiling.  In all cases, the recovered $v\sin i$
was greater than or equal to but always within 5 km s$^{-1}$ of the 25
km s$^{-1}$ we put into the spectra.

Table 2 also lists the values of r$_{\rm Na}$ derived for the MK
standards to which we had artificially added an r$_{\rm Na}$=2.0.  As
in the case of T$_{\rm eff}$, the fit for HD~117176 differs strongly
from the input value.  For the remaining luminosity class V sources,
the derived r$_{\rm Na}$ is 2.7 $\pm$0.22.  For the two luminosity
class IV sources, the average value is 2.4.  One possible contributor
to this systematic difference may be the damping parameter used in the
line synthesis. The parameter that works for the solar spectrum may
not be ideal for the cooler target stars.  For the YSOs, whose
gravities are more comparable to the luminosity class IV MK standards,
this systematic problem may lead to an overestimate of r$_{\rm Na}$ by
$\Delta$r$_{\rm Na}$= 0.13$\times$(1+ r$_{\rm Na}$). Differences
between the value of r$_{\rm Na}$ derived from analysis of the Na
interval spectra and values derived by other methods that are smaller
than this $\Delta$r$_{\rm Na}$ are probably not significant. Further
analysis with a better sample of MK standards will be needed to
understand this effect more fully.

On a real sample of PMS stars, one would be forced to address the
effects of possible binary companions.  Close companions will be
recognizable in high resolution spectra because of their effect on
stellar radial velocities.  Beyond a few AU, the radial velocity
effects will be much less apparent and imaging will be needed.  A
recent compilation of multiplicity data yields a companion star
frequency for TTSs in nearby star forming regions of 24\% $\pm$11\%
(Ophiuchus) and 37\% $\pm$9\% (Taurus), \citep{barsony2003}, for
bright companions ($\Delta$K $\le$ 2--3 mag).  Of these, 40\% have
large enough separations to make them readily separable for direct
imaging or spectroscopy.  Therefore, $\sim$20\% of a sample of stars
in nearby star forming regions will have spectra that suffer from
contamination of a secondary component.  Spectroscopy using adaptive
optics will eliminate this problem for the vast majority of sources.

Continuum opacity due to molecules not present in our synthesis models
is also a likely contributor to the systematic difference in the
derived veiling values.  Preliminary tests with numerous CO, SiO, OH,
and H$_2$O lines that blanket the Na interval reveal a $\sim$5\%
decrease in the continuum compared to the continuum determined from
our synthesis models (Carbon 2003, private communication).  Neglecting
this decrease in the continuum relative to the cores of the
photospheric lines could cause overestimates in the derived veiling by
$\sim$0.05, as well as slightly altering the shapes of the atomic line
wings.

\section{Conclusions}

We have described, demonstrated and evaluated a technique for deriving
effective temperatures, surface gravities, rotation rates, and
infrared excesses from high resolution spectra of PMS stars in the
near--IR.  In a companion paper (DJW03), we use this technique to
study a sample of YSOs in the $\rho$ Ophiuchi molecular cloud core.
Using the Na interval at 2.2~$\mu$m, we can recover the effective
temperatures of dwarf MK standards at a level below the uncertainty in
the spectral type--temperature conversion.  The spectra also give us a
good measure of the rotation velocity and continuum veiling.  With the
addition of a measurement of the relative flux between the
(2--0)~$^{12}$CO bandhead and the Na interval, it is possible to
determine the surface gravity of YSOs and to remove uncertainties in
the temperature caused by a partial degeneracy with $\log g$. The
derived parameters are insensitive to extinction along the
line--of--sight and need photometric information only to tie together
the intensity scales in non--contiguous high resolution spectra and to
correct for differential reddening between 2.2 $\mu$m and 2.3 $\mu$m.

It is clear from the results that the ability to work with weak
features and to measure line shapes as well as equivalent widths can
make high resolution near--IR spectroscopy a valuable tool for studies
of highly obscured stars and of young objects with strong excess
infrared emission.  The technique we develop here is robust enough to
deal with sources with a range of different S/N ratios and with excess
near--IR emission.

In the future, it will be worthwhile to investigate whether high
resolution spectra of other near--IR intervals could add useful
information about the stellar parameters.  We would also like to
investigate more sophisticated matching routines that might make
better use of the sensitivity of individual weak features to various
stellar parameters.  For studies of cooler, lower mass YSOs, it would
be useful to produce models and line syntheses for lower temperature
objects.

\acknowledgments

We thank Chris Sneden for his advice and support in adapting MOOG for
use with YSOs, Tom Greene for providing us with his data on MK
standards, Kevin Luhman for making available his sample of Ophiuchus
spectra, Ken Hinkle for help with our PHOENIX observations, and Russel
White and Carlos Allende--Prieto for helpful comments.  Early phases
of this work were supported in part by NSF grant AST 95-30695 to the
University of Texas.

\appendix

\section{Surface Gravity Diagnostics}

\begin{figure}
  \epsscale{0.8} \plotone{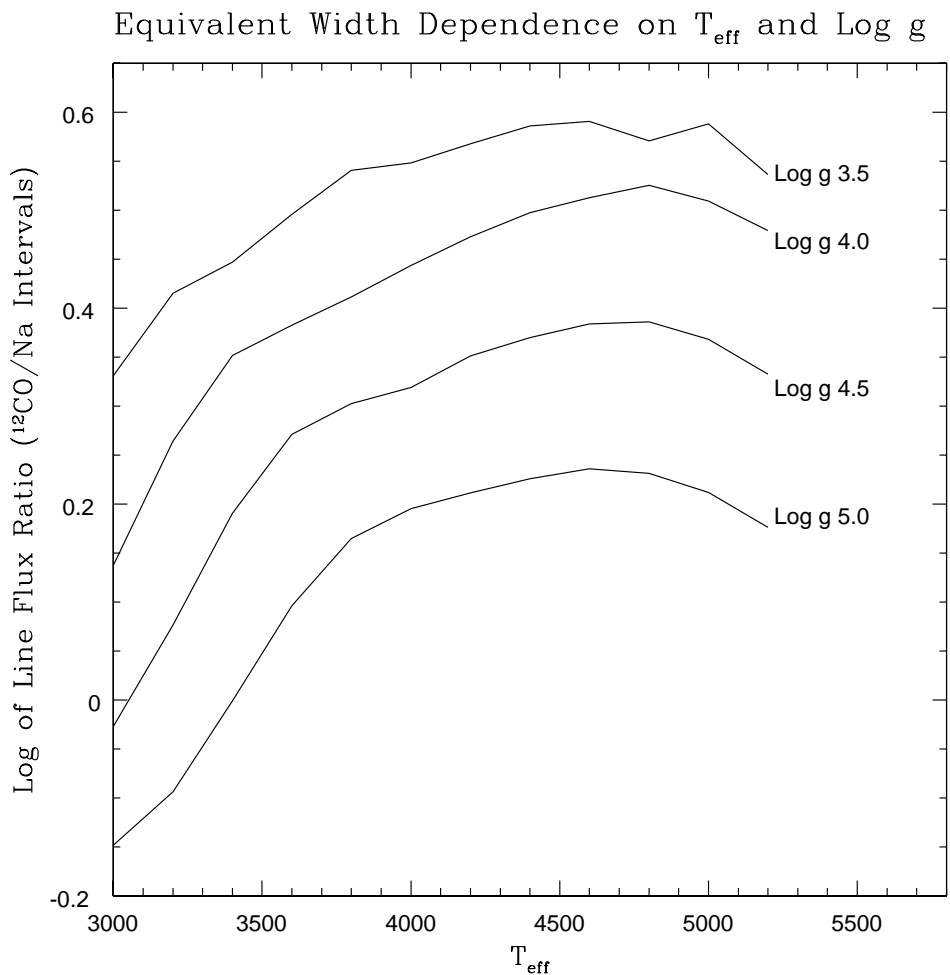}
\caption[The ratio of CO interval equivalent width to Na interval
equivalent width as a function of temperature] 
{\label{fig-isograv} The ratio of CO interval equivalent width to Na
interval equivalent width as a function of temperature, plotted for
surfaces gravities between $\log g=3.5$ and $\log g=5.0$.  Equivalent
widths were computed over the Na and $^{12}$CO intervals as defined in
Figure~\ref{fig-lambda.intervals}.  The relatively flat shape of these
isogravity lines with temperature illustrates the value of the
$^{12}$CO/Na line flux ratio as a good diagnostic of surface gravity.}
\end{figure}

The ratio of photospheric equivalent widths from the Na doublet at
2.2~$\mu$m and the (2--0)~$^{12}$CO bandhead at 2.3~$\mu$m is
sensitive to changes in surface gravity.  Figure~\ref{fig-isograv}
shows that this ratio gives a good estimate of $\log g$ across the
entire range in effective temperature and surface gravity relevant to
the study of low mass PMS stars.  In that figure, the photospheric
equivalent widths in both the Na interval and the $^{12}$CO interval
were calculated from model spectra synthesized from the NextGen
stellar atmosphere models for $\log g=3.5$--5.0 and T$_{\rm
  eff}=3000$--5000~K.

The problem in studying YSOs is that they suffer both from significant
extinction and reddening and often from the presence of excess
continuum emission that can be larger than the emission from the
photospheres themselves.  As a result, the measured equivalent width
ratios do not accurately reflect the ratio of photospheric equivalent
widths that would be relevant for comparison with
Figure~\ref{fig-isograv}.  What we would like to be able to do is to
take the observed equivalent widths and, using as little additional
data as possible and in a way as insensitive as possible to the
effects of reddening and infrared excess, correct the observed
equivalent width ratio to the photospheric equivalent width ratio.  We
outline here a procedure that uses near--IR photometry to correct for
differential reddening between 2.2~$\mu$m and 2.3~$\mu$m and low
resolution spectroscopic data (if necessary) to correct for throughput
differences between the high resolution spectra in the Na and
$^{12}$CO intervals.  In both cases, the corrections are usually quite
small.  To make it easier to use the equivalent width ratio as a
diagnostic for surface gravity, the expressions below are geared to
observable quantities.
 
Starting with an absorption spectrum at high resolution, we define the
measured equivalent width (MEW) in terms of the {\it measured} flux
($F_{\lambda}$) relative to the {\it measured} continuum
($c_{\lambda}$):
 
\begin{equation}
\label{eqn-mew.def}
{\rm MEW}=\int {\frac{c_{\lambda }-F_{\lambda}}{c_{\lambda }}}\,d\lambda
\end{equation}
 
We define the photospheric equivalent width (PEW) to be the equivalent
width of a photospheric absorption line without its value being
altered by the presence of continuum veiling originating outside of
the photosphere (r$_{\lambda}$=(F$_{\rm source}$~--~F$_{\rm
phot}$)/F$_{\rm phot}$).  We define, then, photospheric equivalent
widths for the regions around the Na features at 2.2~$\mu$m (PEW$_{\rm
Na}$) and the shortest wavelength part of the (2--0)~$^{12}$CO
R--branch at 2.3~$\mu$m (PEW$_{\rm CO}$).
 
\begin{equation}
\label{eqn-pew.na}
{\rm PEW}_{\rm Na}=\int_{2.202\mu \rm m}^{2.212\mu \rm m} \frac{c_{2.2\mu 
\rm m}-F(\lambda)}{c_{2.2\mu \rm m}} (1+r_{2.2\mu \rm m})\,d\lambda
\end{equation}

\begin{equation}
\label{eqn-pew.co}
{\rm PEW}_{\rm CO}=\int_{2.2925\mu \rm m}^{2.3000\mu \rm m} \frac{c_{2.3\mu 
\rm m}-F(\lambda)}{c_{2.3\mu \rm m}} (1+r_{2.3\mu \rm m})\,d\lambda
\end{equation}
 
The photospheric continuum $p_{\lambda}$ is altered by infrared excess
(increasing the continuum by a factor ($1+r_\lambda$)), probably due
to thermal emission from a warm surrounding disk, and extinction
($A_{\lambda}$) along the line--of--sight.  The corresponding
expressions for the measured continuum ($c_{\lambda}$) in the two
wavelength regions of interest are:
 
\begin{equation}
\label{eqn-continuum2.2}
c_{2.2\mu \rm m}=p_{2.2\mu \rm m}(1+r_{2.2\mu \rm m})10^{-0.4A_{2.2\mu 
\rm m}}~~{\rm erg}~{\rm s}^{-1}~{\rm cm}^{-2}~\mu \rm m^{-1}
\end{equation}
 
\begin{equation}
\label{eqn-continuum2.3}
c_{2.3\mu \rm m}=p_{2.3\mu \rm m}(1+r_{2.3\mu \rm m})10^{-0.4A_{2.3\mu 
\rm m}}~~\rm erg~s^{-1}~cm^{-2}~\mu m^{-1}
\end{equation}
 
Knowing the effective temperature (T$_{\rm eff}$) allows us to relate
the photospheric continua to each other by assuming a blackbody
dependence ($B(T,\lambda)$).
 
\begin{equation}
\label{eqn-chi}
\chi(T_{\rm eff})=\frac{p_{2.3\mu \rm m}}{p_{2.2\mu \rm m}}=\frac{B_{\lambda}(T_{\rm eff},2.3\mu 
\rm m)}{B_{\lambda}(T_{\rm eff},2.2\mu \rm m)}
\end{equation}
 
By taking the ratio of
equations~\ref{eqn-continuum2.2}~\&~\ref{eqn-continuum2.3} and
substitution of equation~\ref{eqn-chi}, we arrive at an expression for
a quantity we can measure with low resolution spectra or with cross
dispersed high resolution spectra covering the full range of the two
wavelength intervals: the ratio of measured continua at two different
wavelengths.
 
\begin{equation}
\label{eqn-cont.ratio}
\frac{c_{2.2\mu \rm m}}{c_{2.3\mu \rm m}}=\frac{(1+r_{2.2\mu \rm m})10^{-0.4A_{2.2\mu \rm m}}} 
{\chi(T_{\rm eff})(1+r_{2.3\mu \rm m})10^{-0.4A_{2.3\mu \rm m}}}
\end{equation}

Re--arranging equation \ref{eqn-cont.ratio} and substituting into the
ratio of equations \ref{eqn-pew.na} \& \ref{eqn-pew.co} removes the
dependence on any continuum veiling:

\begin{equation}
\label{eqn-pew.ratio}
\frac{{\rm PEW}_{\rm Na}}{{\rm PEW}_{\rm CO}}=
\frac{{\rm MEW}_{\rm Na}}{{\rm MEW}_{\rm CO}}\frac{c_{2.2\mu 
\rm m}}{c_{2.3\mu \rm m}}\chi(T_{\rm eff})10^{-0.4(A_{2.3\mu \rm m}-A_{2.2\mu \rm m})}
\end{equation}
 
This expression provides terms on the right hand side that we can
measure with spectroscopy and photometry allowing us to utilize the
surface gravity dependence on photospheric equivalent ratios in models
as illustrated in Figure~\ref{fig-isograv}.

% add figures and tables

\clearpage

% Figure 1

\clearpage

% Figure 2

\clearpage

% Figure 3

\clearpage

% Figure 4

\clearpage

% Figure 5

\clearpage

% Figure 6

\clearpage

% Figure 7

\clearpage

% Figure 8

\clearpage
% Figure 9

\clearpage

\begin{deluxetable}{cccccccc}
\footnotesize
\tablecaption{Observed MK standards \label{tbl-mk.stands}}
\tablewidth{\hsize}
\tablehead{
\colhead{Standard} & \colhead{Spectral} & \colhead{Optical} 
&\colhead{Near--IR} & \colhead{Optical} &\colhead{Near--IR} 
&\colhead{Metal-} & S/N\\
\colhead{Star} 
&\colhead{Type}  &\colhead{$T_{\rm eff}$\tablenotemark{c}} &\colhead{$T_{\rm eff}$\tablenotemark{d}} 
&\colhead{$v\sin i$} &\colhead{$v\sin i$} 
&\colhead{licity}\\
&&\colhead{(K)}&\colhead{(K)}&\colhead{(km~s$^{-1})$}&\colhead{(km~s$^{-1})$}&\colhead{[Fe/H]}
}
\startdata

HD 117176\tablenotemark{a}       &G4V\tablenotemark{1}   &5636   &5980   &10\tablenotemark{8}    &$<$3   
&$-$0.11\tablenotemark{3} & 120\\
HR 4496\tablenotemark{b}         &G8V\tablenotemark{1}   &5439   &5400   &$<$15\tablenotemark{8} &$<$9   
&$-$0.14\tablenotemark{4}& 300\\
HR 4496\tablenotemark{a}         &G8V\tablenotemark{1}   &5439   &5460   &$<$15\tablenotemark{8} &$<$3   
&$-$0.14\tablenotemark{4}& 110\\
HD 185144\tablenotemark{a}       &K0V\tablenotemark{1}   &5152   &5260   &$<$15\tablenotemark{8} &$<$3   
&$-$0.23\tablenotemark{3}& 120\\
GL 28\tablenotemark{b}           &K2V\tablenotemark{1}   &4838   &4880   &2.5\tablenotemark{9}   &$<$9   
&$-$0.05\tablenotemark{6}& 180\\
HR 5568\tablenotemark{b}         &K4V\tablenotemark{1}   &4539   &4580   &$<$12\tablenotemark{8} &$<$9   
&0.016\tablenotemark{7}& 170\\
GL 338A\tablenotemark{b}         &M0V\tablenotemark{2}   &3837   &3660   &2.9\tablenotemark{10}   &$<$9   & ...      
& 155\\
GL 338A\tablenotemark{a}         &M0V\tablenotemark{2}   &3837   &3650   &2.9\tablenotemark{10}   &$<$3   &  ...     
& 110\\
HD 131976\tablenotemark{a}       &M1.5V\tablenotemark{1} &3589   &3610   &1.4\tablenotemark{11}   &10     &  ...     
& 75\\
GL 15A\tablenotemark{b}          &M2V\tablenotemark{1}   &3523   &3630   &$<$2.9\tablenotemark{10}        &$<$9   &  ...     
& 150\\
GL 402\tablenotemark{b}          &M4V\tablenotemark{2}   &3289   &3290   &$<$2.3\tablenotemark{10}        &$<$9   &  ...     
& 110\\
%GL 406\tablenotemark{b}         &M6V\tablenotemark{2}   &3031   &3330   &$<$2.9\tablenotemark{10}        &30     & \nl

HD 188512\tablenotemark{a}       &G8IV\tablenotemark{1}  &5100\tablenotemark{5}  &4700   
&1.8\tablenotemark{13}   &$<$3   &$-$0.30\tablenotemark{5}& 130\\
HD 142091\tablenotemark{a}       &K1IVa\tablenotemark{1} &4800\tablenotemark{5}  &4680   
&1.9\tablenotemark{14}   &$<$3   &$-$0.04\tablenotemark{5}& 100\\

\enddata

\tablenotetext{a}{Observations made using PHOENIX on the KPNO 4--meter}
\tablenotetext{b}{Observations made using NIRSPEC on Keck}
\tablenotetext{c}{Using $T_{\rm eff}$--spectral type relation of \citet{dejager1987} 
for luminosity class V sources unless otherwise noted}
\tablenotetext{d}{Assumed $\log g$=4.5 for dwarfs and $\log g$=3.5 for the two 
sub--giants sources \citep{mcwilliam1990}}
\tablerefs{(1) \citet{keenan1989} (2) \citet{kirkpatrick1991} (3) \citet{hearnshaw1974b} (4) \citet{hearnshaw1974a} (5) \citet{mcwilliam1990} (6) \citet{marsakov1988} (7) \citet{taylor1995} (8)\citet{glebocki2000} (9) \citet{strassmeier2000} (10) \citet{delfosse1998} (11) \citet{duquennoy1988} (12) \citet{vogt1983} (13) \citet{demedeiros1999} (14) \citet{fekel1997}}
 
\end{deluxetable}

\clearpage

% Table 2

\begin{deluxetable}{cccccc}
\footnotesize
\tablecaption{Recovered parameters with rotation ($v\sin i =25$~{\rm km~s}$^{-1}$) and 
veiling (r$_{\rm Na}=2.0$) added to MK standards \label{tbl-mkstands.veiling}}
\tablewidth{5.5in}
\tablehead{
\colhead{MK} & \colhead{Best Fit} & \colhead{Recovered} & \colhead{Recovered} 
& \colhead{Recovered} & \colhead{RMS} \\ \colhead{Standard} & 
\colhead{T$_{\rm eff}$} & \colhead{T$_{\rm eff}$} & \colhead{$v\sin i$} &
 \colhead{Veiling} &\\
&\colhead{(K)}&\colhead{(K)}&\colhead{(km~s$^{-1}$)}&\colhead{(r$_{\rm Na}$)}&
}
\startdata

HD 117176       &5980   &5240   &27.0   &3.8    &0.00271\\
HR 4496         &5400   &5230   &29.0   &2.6    &0.00152\\
HD 185144       &5240   &5000   &29.0   &2.7    &0.00189\\
GL 338A         &3650   &3765   &25.0   &3.0    &0.00333\\
HD 131976       &3610   &3580   &30.0   &2.5    &0.00244\\
HD 188512       &4700   &4620   &30.0   &2.3    &0.00165\\
HD 142091       &4680   &4620   &29.0   &2.5    &0.00209\\

\enddata
\end{deluxetable}


\begin{thebibliography}{}
  
\bibitem[Ali et al.(1995)] {ali1995} Ali, B., Carr, J.S., DePoy, D.L., Frogel, J.A., \& Sellgren, K. 1995, \aj, 110, 2415
\bibitem[Allen(2000)]{allen2000} Allen, C. W. 2000, {\em Allen's Astrophyical Quantities} (New York, Springer--Verlag)
\bibitem[Alonso et al.(1996)Alonso, Arribas, \& Martinez--Roger]{alonso1996} Alonso, A., Arribas, S., \& Martinez--Roger, C.\ 1996, \aap, 313, 873
\bibitem[Baldwin et al.(1973)Baldwin, Frogel, \& Persson]{baldwin1973} Baldwin, J.R., Frogel, J.A., \& Persson, S.E. 1973, \apj, 184, 427
\bibitem[Baraffe et al.(1998)Baraffe, Chabrier, Allard, \& Hauschildt]{baraffe1998} Baraffe, I., Chabrier, G., Allard, F., \& Hauschildt, P.~H.\ 1998, \aap, 337, 403
\bibitem[Barsony et al.(1997)Barsony, Kenyon, Lada, \& Teuben]{barsony1997} Barsony, M., Kenyon, S.J., Lada, E.A., \& Teuben P.J. 1997, \apjs, 112, 109
\bibitem[Barsony, Koresko, \& Matthews(2003)]{barsony2003} Barsony, M., Koresko, C., \& Matthews, K.\ 2003, \apj, 591, 1064
\bibitem[Bontemps et al.(2001)]{bontemps2001} Bontemps, S.~et al.\ 2001, \aap,   372, 173
\bibitem[Casali~\& Matthews(1992)]{casali1992} Casali, M.~M.~\& Matthews, H.~E.\ 1992, \mnras, 258, 399
\bibitem[Casali \& Eiroa(1996)]{casali1996} Casali, M.~M.~\& Eiroa, C.\ 1996, \aap, 306, 427
\bibitem[Comer\'{o}n et al(1993)Comer\'{o}n, Rieke, Burrows, \& Rieke]{comeron1993} Comer\'{o}n, F., Rieke, G.~H., Burrows, A., \& Rieke, M.~J.\ 1993, \apj, 416, 185
\bibitem[D'Antona \& Mazzitelli(1997)]{dantona1997} D'Antona, F.~\& Mazzitelli, I.\ 1997, Mem. Soc. Astron. Italiana, 68, 823
\bibitem[Delfosse et al.(1998)Delfosse, Forveille, Perrier, \& Mayor]{delfosse1998} Delfosse, X., Forveille, T., Perrier, C., \& Mayor, M.\ 1998, \aap, 331, 581
\bibitem[de Jager \& Nieuwenhuijzen(1987)]{dejager1987} de Jager, C.~\& Nieuwenhuijzen, H.\ 1987, \aap, 177, 217
\bibitem[de Medeiros \& Mayor(1999)]{demedeiros1999} de Medeiros, J.~R.~\& Mayor, M.\ 1999, VizieR Online Data Catalog, 413, 990433
\bibitem[Doppmann et al.(2003)Doppmann, Jaffe, \& White]{doppmann2002b} Doppmann, G.W., Jaffe, D.T., White, R.J. 2003 \aj, submitted (DJW03)
\bibitem[Duquennoy \& Mayor(1988)]{duquennoy1988} Duquennoy, A.~\& Mayor, M.\ 1988, \aap, 200, 135
\bibitem[Duquennoy \& Mayor(1991)]{duquennoy1991} Duquennoy, A.~\& Mayor, M.\ 1991, \aap, 248, 485
\bibitem[Edvardsson et al.(1993)]{edvardsson1993} Edvardsson, B., Andersen, J., Gustafsson, B., Lambert, D.~L., Nissen, P.~E., \& Tomkin, J.\ 1993, \aap, 275, 101
\bibitem[Fekel(1997)]{fekel1997} Fekel, F.C. 1997, PASP, 109, 514
\bibitem[Ghez et al.(1993)Ghez, Neugebauer, \& Matthews]{ghez1993} Ghez, A.~M., Neugebauer, G., \& Matthews, K.\ 1993, \aj, 106, 2005
\bibitem[Glebocki \& Stawikowski(2000)]{glebocki2000} Glebocki, R.~\& Stawikowski, A.\ 2000, Acta Astronomica, 50, 509
\bibitem[Goorvitch \& Chackerian(1994)]{goorvitch1994} Goorvitch, D. \& Chackerian, C. Jr. 1994, \apjs, 91, 483
\bibitem[Gray(1992)]{gray1992} Gray, D.~F. 1992, The Observation and Analysis of Stellar Photospheres, (New York, Cambridge Univ. Press)
\bibitem[Gray, Graham, \& Hoyt(2001)]{gray2001} Gray, R.~O.,
Graham, P.~W., \& Hoyt, S.~R.\ 2001, \aj, 121, 2159
\bibitem[Greene~\& Young(1992)]{greene1992} Greene, T.P., \& Young, E.T. 1992, \apj, 395, 516
\bibitem[Greene et al.(1993)Greene, Tokunaga, Toomey, \& Carr]{greene1993} Greene, T.~P., Tokunaga, A.~T., Toomey, D.~W., \& Carr, J.~B.\ 1993, \procspie, 1946, 313
\bibitem[Greene et al.(1994)]{GWAYL1994} Greene, T.~P., Wilking, B.~A., Andr\'{e}, P., Young, E.~T., \& Lada, C.~J.\ 1994, \apj, 434, 614
\bibitem[Greene \& Meyer(1995)]{greene1995} Greene, T.~P.~\& Meyer, M.~R.\ 1995, \apj, 450, 233
\bibitem[Greene~\& Lada(1996)]{greene1996} Greene, T.~P.~\& Lada, C.~J.\ 1996, \aj, 112, 2184
\bibitem[Greene \& Lada(1997)]{greene1997} Greene, T. P. \& Lada, C. J. 1997, \apj, 114, 2157
\bibitem[Greene \& Lada(2000)]{greene2000} Greene, T. P. \& Lada, C. J. 2000, \apj, 120, 430
\bibitem[Greene \& Lada(2002)]{greene2002} Greene, T.~P.~\& Lada, C.~J.\ 2002, \aj, 124, 2185
\bibitem[Gullbring et al.(2000)Gullbring, Calvet, Muzerolle, \& Hartmann]{gullbring2000} Gullbring, E., Calvet, N., Muzerolle, J., \& Hartmann, L.\ 2000, \apj, 544, 927
\bibitem[Hatmann et al.(1991)Hartmann, Stauffer, Kenyon, \& Jones(1991)]{hartmann1991} Hartmann, L., Stauffer, J.~R., Kenyon, S.~J., \& Jones, B.~F.\ 1991, \aj, 101, 1050
\bibitem[Hauschildt et al.(1999)Hauschildt, Allard, \& Baron]{hauschildt1999} Hauschildt P.H., Allard, F. \& Baron, E. 1999, \apj, 512, 377
\bibitem[Hearnshaw(1974a)]{hearnshaw1974a} Hearnshaw, J.~B.\ 1974, \aap, 34, 263
\bibitem[Hearnshaw(1974b)]{hearnshaw1974b} Hearnshaw, J.~B.\ 1974, \aap, 36, 191
\bibitem[Hinkle et al.(1998)]{hinkle1998} Hinkle, K.~H., Cuberly, R.~W., Gaughan, N.~A., Heynssens, J.~B., Joyce, R.~R., Ridgway, S.~T., Schmitt, P., \& Simmons, J.~E.\ 1998, \procspie, 3354, 810
\bibitem[Huang(1961)]{huang1961} Huang, S.\ 1961, \apj, 134, 12
\bibitem[Johns--Krull \& Valenti(2001)]{johnskrull2001} Johns--Krull, C.~M.~\& Valenti, J.~A.\ 2001, \apj, 561, 1060
\bibitem[Keenan \& McNeil(1989)]{keenan1989} Keenan, P.~C.~\& McNeil, R.~C.\ 1989, \apjs, 71, 245
\bibitem[Kenyon \& Hartmann(1995)]{kenyon1995} Kenyon, S.~J.~\& Hartmann, L.\ 1995, \apjs, 101, 117
\bibitem[Kenyon et al.(1998)Kenyon, Brown, Tout, \& Berlind]{kenyon1998} Kenyon, S.~J., Brown, D.~I., Tout, C.~A., \& Berlind, P.\ 1998, \apj, 115, 2491
\bibitem[Kirkpatrick et al.(1991)Kirkpatrick, Henry, \& McCarthy]{kirkpatrick1991} Kirkpatrick, J.~D., Henry, T.~J., \& McCarthy, D.~W.\ 1991, \apjs, 77, 417
\bibitem[Kirkpatrick et al.(1993)]{kirkpatrick1993} Kirkpatrick, J.D., Kelly, D.M., Rieke, G.H., Liebert, J. Allard, F., \& Wehrse, R. 1993, \apj, 402, 643
\bibitem[Kleinmann \& Hall(1986)] {kleinmann1986} Kleinmann, S. G., \& Hall, D. N. B. 1986, \apjs, 62, 501
\bibitem[Kurucz(1994)]{kurucz1994} Kurucz, R. L. 1994, Atomic Data for Opacity Calculations, Kurucz CD--ROM No. 1
\bibitem[Lada(1987)]{lada1987} Lada, C.~J.\ 1987, IAU Symp.~115: Star Forming Regions, 115, 1
\bibitem[Lancon \& Rocca--Volmerange(1992)] {lancon1992} Lancon, A., \& Rocca--Volmerange, B. 1992, \aaps, 96, 593
\bibitem[Leggett et al.(1996)]{leggett1996} Leggett, S.K., Allard, F., Berriman, G., Dahn, C.C., and Hauschildt, P.H. 1996, \apjs, 104, 117
\bibitem[Livingston \& Wallace(1991)]{livingston1991} Livingston, W. \& Wallace L. 1991 An Atlas of the Solar Spectrum in the Infrared from 1850 to 9000 cm-1 (1.1 to 5.4 microns), N.S.O. Technical Report \#91-001, July 1991
\bibitem[Luhman \& Rieke(1999)]{luhman1999} Luhman. K. L. \& Rieke, G. H. 1999, \apj, 525, 440
\bibitem[Marsakov \& Shevelev(1988)]{marsakov1988} Marsakov, V.~A.~\& Shevelev, Y.~G.\ 1988, Bulletin d'Information du Centre de Donnees Stellaires, 35, 129
\bibitem[McCaughrean(2001)]{mccaughrean2001} McCaughrean, M.~J.\ 2001, IAU Symposium, 200, 169
\bibitem[McLean et al.(1995)]{mclean1998} McLean, I.~S., Becklin, E.~E., Figer, D.~F., Larson, S., Liu, T., \& Graham, J.\ 1995, \procspie, 2475, 350
\bibitem[McWilliam(1990)]{mcwilliam1990} McWilliam, A.\ 1990, \apjs, 74, 1075
\bibitem[Merrill \& Ridgway(1979)] {merrill1979} Merrill, K.M., \& Ridgway, S.T. 1979, \araa, 17, 9
\bibitem[Meyer et al.(1998)Meyer, Edwards, Hinkle, \& Strom]{meyer1998} Meyer, M. R., Edwards, S., Hinkle, K.\ H., \& Strom, S.\ E.\ 1998, \apj, 508, 397
\bibitem[Mountain et al.(1990)]{mountain1990} Mountain, C.M., Robertson, D.J., \ Lee, T.J., \& Wade, R. 1990, in Instrumentation in Astronomy, \procspie, p. 25
\bibitem[Padgett(1996)]{padgett1996} Padgett, D.~L.\ 1996, \apj, 471, 847
\bibitem[Palla \& Stahler(1999)]{palla1999} Palla, F.~\& Stahler, S.~W.\ 1999, \apj, 525, 772
\bibitem[Palla \& Stahler(2000)]{palla2000} Palla, F.~\& Stahler, S.~W.\ 2000,   \apj, 540, 255
\bibitem[Ram\'{i}rez et al.(1997)]{ramirez1997} Ram\'{i}rez, S.V., DePoy, D.L., Frogel, J.A., Sellgren, K., \& Blum, R.D. 1997, \aj, 113, 1411
\bibitem[Siess et al.(2000)Siess, Dufour, \& Forestini]{siess2000} Siess, L., Dufour, E., \& Forestini, M. 2000, \aap, 358, 593
\bibitem[Simon et al.(1995)]{simon1995} Simon, M.~et al.\ 1995, \apj, 443, 625
\bibitem[Simon et al.(2000)]{simon2000} Simon, M., Dutrey, A., \&  Guilloteau, S. 2000, ApJ, 545, 1034
\bibitem[Sneden(1973)]{sneden1973} Sneden, C. 1973, Ph.D. Thesis, University of Texas at Austin
\bibitem[Stahler(1988)]{stahler1988} Stahler, S.~W.\ 1988, \pasp, 100, 1474
\bibitem[Strassmeier et al.(2000)Strassmeier, Granzer, Scheck, \& Weber]{strassmeier2000} Strassmeier, K.~W.~A., Granzer, T., Scheck, M., \& Weber, M.\ 2000, \aaps, 142, 275
\bibitem[Strom et al.(1995)Strom, Kepner, \& Strom]{strom1995} Strom, K.~M., Kepner, J., \& Strom, S.~E.\ 1995, \apj, 438, 813
\bibitem[Taylor(1995)]{taylor1995} Taylor, B.~J.\ 1995, \pasp, 107, 734
\bibitem[Uns\"{o}ld(1950)]{unsold1955} Uns\"{o}ld, A. 1955, Physik der Sternatmosph\"{a}ren (2nd ed.; Berlin: Springer--Verlag)
\bibitem[Vogt et al.(1983)Vogt, Penrod, \& Soderblom]{vogt1983} Vogt, S.~S., Penrod, G.~D., \& Soderblom, D.~R.\ 1983, \apj, 269, 250\bibitem[Wallace \& Hinkle(1996)]{wallace1996} Wallace, L., \& Hinkle, K. 1996, \apjs, 107, 312
\bibitem[Webb et al.(1999)]{webb1999} Webb, R.~A., Zuckerman, B., Platais, I., Patience, J., White, R.~J., Schwartz, M.~J., \& McCarthy, C.\ 1999, \apjl, 512, L63
\bibitem[White et al.(1999)White, Ghez, Reid, \& Schultz]{white1999} White, R.~J., Ghez, A.~M., Reid, I.~N., \& Schultz, G.\ 1999, \apj, 520, 811
\bibitem[White \& Ghez(2001)]{white2001} White, R.~J.~\& Ghez, A.~M.\ 2001, \apj, 556, 265
\bibitem[Wilking~\& Lada(1983)]{wilking1983} Wilking, B.~A., \& Lada, C.~J. 1983, \apj, 274, 698
\bibitem[Wilking et al.(1989)Wilking, Lada, \& Young]{wilking1989} Wilking, B.~A., Lada, C.~J., \& Young, E.~T.\ 1989, \apj, 340, 823
\bibitem[Wright et al.(1993)]{wright1993} Wright, G.~S., Mountain, C.~M., Bridger, A., Daly, P.~N., Griffin, J.~L., \& Ramsay Howat, S.~K.\ 1993, \procspie, 1946, 547

\end{thebibliography}
\end{document}